\documentclass[]{emulateapj}

\usepackage{natbib}
\usepackage{lscape}
\usepackage{color}

\newcommand{\toneC}{$\rm \theta^1C$}
\newcommand{\toneCone}{$\rm \theta^1C_1$}
\newcommand{\toneCtwo}{$\rm \theta^1C_2$}
\newcommand{\caseone}{$\rm [\theta^1C_1,\theta^1C_2]\:BN \longrightarrow [\theta^1C_1,\theta^1C_2]\: BN$}
\newcommand{\casetwo}{$\rm [\theta^1C_1,BN]\:\theta^1C_2 \longrightarrow [\theta^1C_1,\theta^1C_2]\: BN$}
\newcommand{\casethree}{$\rm [\theta^1C_2 , BN]\:\theta^1C_1 \longrightarrow [\theta^1C_1,\theta^1C_2]\: BN$}

\newcommand{\msun}{\rm{M_\odot}}
\newcommand{\lsun}{\rm{L_\odot}}
\newcommand{\au}{\rm{AU}}

\newcommand{\code}[1]{{\tt #1}}

\newcommand{\kms}{\rm km\:s^{-1}}
\newcommand{\truebn}{{\tt BN-True}}
\newcommand{\bnvel}{{\tt BN-Velocity}}
\newcommand{\bnejection}{{\tt BN-Ejection}}

\shorttitle{Runaway star via binary-single interactions}
\shortauthors{Chatterjee et al.}

\begin{document}


\title{Gravitational slingshot of young massive stars in Orion}


\author{Sourav\,Chatterjee\altaffilmark{1}}
\email{s.chatterjee@astro.ufl.edu}
\author{Jonathan~C.~Tan\altaffilmark{1,2}}
\email{jt@astro.ufl.edu}

\altaffiltext{1}{Department of Astronomy, University of Florida, Gainesville, FL 32611.  }
\altaffiltext{2}{Department of Physics, University of Florida, Gainesville, FL 32611.  }

\date{}

\begin{abstract}
The Orion Nebula Cluster (ONC) is the nearest region of massive star
formation and thus a crucial testing ground for theoretical models. Of
particular interest amongst the ONC's $\sim 1000$ members are:
$\theta^1$~Ori~C, the most massive binary in the cluster with stars of
masses $38$ and $9\,\msun$ \citep{2009A&A...497..195K}; the
Becklin-Neugebauer (BN) object, a $30\,\kms$ runaway star of $\sim
8\,\msun$ \citep{2004ApJ...607L..47T}; and the Kleinmann-Low (KL)
nebula protostar, a highly-obscured, $\sim 15\,\msun$ object still
accreting gas while also driving a powerful, apparently ``explosive''
outflow \citep{1993Natur.363...54A}.  The unusual behavior of BN and
KL is much debated: How did BN acquire its high velocity?  How is this
related to massive star formation in the KL nebula? Here we report the
results of a systematic survey using $\sim 10^7$ numerical experiments
of gravitational interactions of the \toneC\ and BN stars.  We show
that dynamical ejection of BN from this triple system at its observed
velocity leaves behind a binary with total energy and eccentricity
matching those observed for \toneC. Five other observed properties of
\toneC\ are also consistent with it having ejected BN and altogether
we estimate there is only a $\lesssim 10^{-5}$ probability that
\toneC\ has these properties by chance. We conclude that BN was
dynamically ejected from the \toneC\ system about 4,500 years ago. BN
has then plowed through the KL massive-star-forming core within the
last 1,000 years causing its recently-enhanced accretion and outflow
activity.
\end{abstract}





\section{Introduction}
\label{sec:intro}
Massive stars impact many areas of astrophysics. In most galactic 
environments they dominate the radiative, mechanical and chemical 
feedback on the interstellar medium, thus regulating the evolution of 
galaxies. Many low-mass stars form in clusters near massive stars, and 
their protoplanetary disks can be affected by this feedback also. There is 
some evidence that our own solar system was influenced in this way 
\citep[e.g.,][]{2003ApJ...588L..41T,2010ARA&A..48...47A}. 
Despite this importance, there is no consensus 
on the basic formation mechanism of massive stars. Theories range from 
scaled-up versions of low-mass star formation \citep{2003ApJ...585..850M}, to 
competitive Bondi-Hoyle accretion at the center of forming star clusters 
\citep{2001MNRAS.323..785B,2010ApJ...709...27W}, to stellar collisions 
\citep{1998MNRAS.298...93B}.
The Orion Nebula Cluster (ONC) is the nearest region of massive star
formation and thus a crucial testing ground for theoretical models. 

Of particular interest amongst the ONC's $\sim 1000$ members in this regard are:
$\theta^1$~Ori~C, the most massive binary in the cluster with stars of
masses $\approx 38$ and $9\,\msun$ \citep{2009A&A...497..195K}; the
Becklin-Neugebauer (BN) object, a $30\,\kms$ runaway star of
$\approx 8 M_\odot$ \citep{2004ApJ...607L..47T}; and the
Kleinmann-Low (KL) nebula protostar, a highly-obscured, about $15\,\msun$ 
object still accreting gas while also driving a
powerful, apparently ``explosive'' outflow \citep{1993Natur.363...54A}. 
The unusual behavior of BN and KL is much debated bearing implications 
towards massive-star formation theories: How did BN acquire its high velocity?  How is this
related to massive star formation in the KL nebula? 

BN, like KL, is heavily obscured by dust so its luminosity of $\sim
(5\pm3)\times 10^3\,\lsun$ mostly emerges in the
infrared \citep{1998ApJ...509..283G}. The above luminosity constrains
BN's mass to be $m_{\rm BN} \simeq 9.3\pm2.0\msun$, assuming it is on
the zero age main sequence \citep{2004ApJ...607L..47T}.  For this
estimate and throughout the paper we have adopted $414\pm7$~pc for the
distance to the cluster \citep{2007A&A...474..515M}.  Astrometry based
on mm and radio observations indicate that BN is a runaway
star \citep{1995ApJ...455L.189P,2004ApJ...607L..47T}, with some recent
measurements of its motion in the ONC frame of $\mu_{\rm BN}=13.2\pm
1.1\:{\rm mas\:yr^{-1}}$ towards P.A.$_{\rm
  BN}=-27^\circ.5\pm4^\circ$\citep{2008ApJ...685..333G} and $\mu_{\rm
  BN}=13.4\pm 1.1\:{\rm mas\:yr^{-1}}$ towards P.A.$_{\rm
  BN}=-18^\circ.8\pm4.6^\circ$ \citep{2011ApJ...728...15G} (Figure\ 1).
This corresponds to a velocity $v_{\rm
  2D,BN}=25.9\pm2.2\:{\rm km\:s^{-1}}$ \citep{2008ApJ...685..333G}. 
  BN has an observed radial
(LSR) velocity of $+21\pm\sim 1\:\kms$ \citep{1983ApJ...275..201S},
while the ONC mean is $+8.0\:\kms$ \citep[based on a mean heliocentric
velocity of about 1000 ONC stars of
$+26.1\:\kms$;][]{2008ApJ...676.1109F}. Including this $+13\pm\sim
1\:{\rm km\:s^{-1}}$ radial velocity with respect to the ONC mean, the
3D ONC-frame velocity of BN is $v_{\rm 3D,BN}=29\pm3\:{\rm
  km\:s^{-1}}$. This is much greater than the velocity dispersion of
ONC stars, variously inferred to be $\sigma_{\rm 3D} = 2.4 \:\kms$
based on the proper motions of $\sim 50$ bright ($V\lesssim 12.5$)
stars within 30$^\prime$ of the ONC center \citep{1988AJ.....95.1744V},
$\sigma_{\rm 3D} = 3.8 \:\kms$ based on proper motions of $\sim 900$
fainter stars within 15$^\prime$ of the ONC
center \citep{1988AJ.....95.1755J}, and $\sigma_{\rm 3D} = 5.4 \:\kms$
based on radial velocity measurements (potentially affected by motion
induced by binarity) of $1100$ stars within $\sim$60$^\prime$ of the
ONC center \citep{2008ApJ...676.1109F}. Thus there is little doubt that BN
is a runaway star, which formed and was then accelerated in the ONC.

Supernova explosion of one member of a binary can lead to the other
being ejected at high speeds \citep{1957moas.book.....Z}. The ONC is too
young \citep[most stars are $<3$~Myr old;][]{2010ApJ...722.1092D} for a
supernova to have occurred. Nor is there any evidence for a recent
supernova.  Alternatively, runaway stars can be produced via dynamical
ejection --- a gravitational slingshot --- from a triple or higher multiple
system \citep{1967BOTT....4...86P,1983ApJ...268..319H,1986ApJS...61..419G}, 
in which the lowest mass member
tends to be ejected. 
Indeed, such dynamical ejection of stars naturally happen in dense and young 
clusters \citep[e.g.,][]{2004MNRAS.350..615G,2011Sci...334.1380F}.  Similar ejections 
have also been discussed in the context of some other ONC stars \citep[][]{2004MNRAS.350..615G}.       
Thus BN, having formed in the ONC, should have
been accelerated via dynamical ejection. 
The predictions of this scenario are very specific:
somewhere along BN's past trajectory should be a massive binary (or
higher order multiple), with two components likely more massive than BN, recoiling
in the opposite direction, and, as we shall see, with specific orbital
properties.

Two scenarios for the dynamical ejection of BN have been
proposed. {\tt (1)} Ejection from the \toneC\ binary
\citep{2004ApJ...607L..47T}: In this scenario the BN star is ejected
via a strong gravitational scattering interaction \citep[e.g.,
][]{1983ApJ...268..319H,2011MNRAS.410..304G} and later plows, {\it by
  chance}, through the KL star-forming core to drive tidally-enhanced
accretion and thus outflow activity. If so, a model of formation of
the KL massive protostar via an ordered collapse of a gas core to a
central disk\citep{2003ApJ...585..850M}, similar to how low-mass stars
are thought to form, is still broadly applicable, though subject to
the tidal perturbation from BN's fly-by. {\tt (2)} Ejection from the
KL (source {\it I}) protostar
\citep{2005AJ....129.2281B,2008ApJ...685..333G}: Here it is proposed
that the KL outflow is related to the disintegration of a {\it
  forming} triple system, which ejected BN and produced a binary
suggested to be the radio source I \citep{1995ApJ...445L.157M}. This binary
has recoiled southwards from the original formation site and is now
hidden, {\it by chance}, behind or in the dense gas core near the
center of the KL nebula. In this scenario the gas from the original
formation site, like the stars, has also been expelled in this event
to form the outflow, and the core that formed these massive stars has
been destroyed.  If true, this is a very different formation process,
and would indicate that chaotic gravitational interactions between
multiple protostars followed by complete ejection of both stars and
gas are intrinsic features of massive star formation
\citep{2005AJ....129.2281B,2011ApJ...727..113B}, at least in this case
in Orion.

Figure~1 shows a near-IR image of the central region of the ONC,
including BN, the KL protostar (marked by radio source {\it
  I}) and the famous Trapezium stars, of
which \toneC\ is the brightest. The past trajectory of BN is indicated based on its
present motion and assuming no acceleration. It goes near
KL {\it and} the Trapezium stars. 
The high obscuration to KL means that there is
little direct constraint on the properties of the star(s): for example
there is no evidence that it is even a binary. In contrast, the
properties of \toneC\ have been measured much more
precisely \citep{2009A&A...497..195K} and so the scenario of ejection 
of BN via a binary-single strong scattering can be tested
much more rigorously and is the goal of this study. 

In \S\ref{sec:method} we summarize our methods and numerical
calculations. In \S\ref{sec:results} we present our key results, which
show that \toneC\ has orbital properties expected if it ejected BN. In
\S\ref{sec:probability} we estimate the probability that \toneC\ has
not been responsible for BN's ejection and has these and other
observed properties simply by chance. We summarize and conclude in
\S\ref{sec:conclude}.

\begin{figure*}
\begin{center}
\plotone{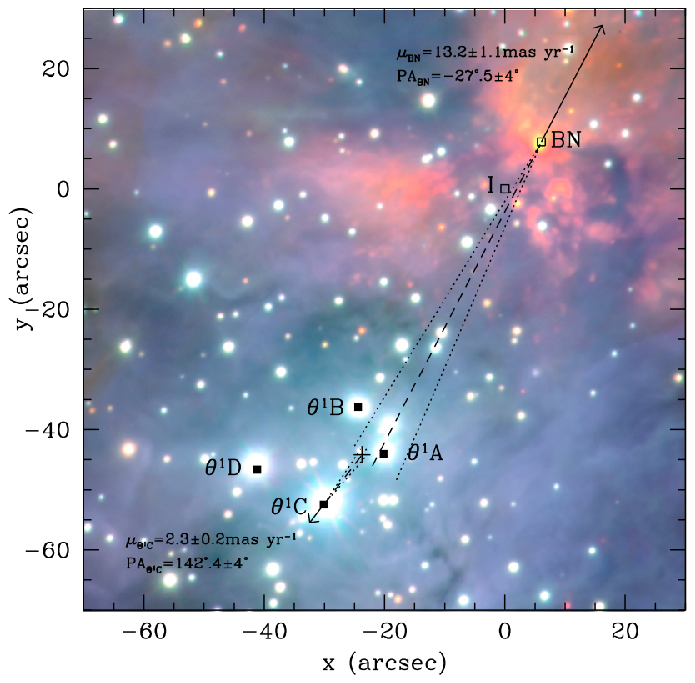} 
\caption[ONC with BN]{\footnotesize Near-infrared (J,H,K) image of the central region of the Orion Nebula
Cluster \citep{2002Msngr.109...28M} with cluster frame proper motions of
BN  \citep{2008ApJ...685..333G,2008arXiv0807.3771T} and
\toneC\  \citep{1988AJ.....95.1744V} indicated by arrows proportional to
the size of the motion. The coordinates are relative to the position
of source {\it I} ($\alpha$(J2000)=05 35 14.5141, $\delta$(J2000)=-05 22
30.556) \citep{2008ApJ...685..333G}. Under the simplifying assumption of
no acceleration, these motions are traced back with dashed lines
(dotted lines indicate $1\sigma$ uncertainties) to a common origin
about 4,500 years ago, shown by the cross at (-23.72, -44.16). The
positions of the other Trapezium stars and the Kleinmann-Low massive
protostar, source {\it I}, are also indicated.}
\label{fig:onc}
\end{center}
\end{figure*}

\section{Methods}
\label{sec:method}
We investigate the scenario of ejection of BN from the \toneC\ binary
by carrying out calculations of gravitational scattering between the
three stars. We adopt the central values of the observed masses of the
\toneC\ binary members \citep{2009A&A...497..195K}, $m_{\rm
  \theta^1C_1}=38.2\pm3.6\,\msun$ and $m_{\rm \theta^1C_2}=8.8\pm
1.7\,\msun$. For BN we adopt the central value of the mass estimate
$m_{\rm BN}=8.2\pm2.8\,\msun$, which is based on the observed
cluster-frame proper motion of \toneC\ of $\mu_{\rm
  \theta^1C}=2.3\pm0.2\:{\rm mas\:yr^{-1}}$
 \citep{1988AJ.....95.1744V}, i.e., $v_{\rm 2D,\theta^1C} =
4.5\pm0.4\:{\rm km\:s^{-1}}$, and assuming it is due to recoil from
ejecting BN. Note that this mass is consistent with that inferred from
the luminosity of BN, discussed above. The error range includes an
assumed pre-ejection motion of the center of mass of the 3 stars along
the ejection axis of 0.7~${\rm mas\:yr^{-1}}$, i.e., similar to that
observed for other bright ONC stars \citep{1988AJ.....95.1744V}.

We carry out a systematic investigation of the three possible strong
scattering interactions between a binary and a single star.  Depending on 
the initial perturber and the binary members these interactions can be divided 
into three types.  {\bf Type 1 - BN star is the perturber:} Here BN is initially a single star 
that interacts with \toneC\ members \toneCone\ and \toneCtwo.  We denote this 
initial configuration as [\toneCone, \toneCtwo]~BN.  The square 
brackets denote a bound binary and the star outside the brackets is a single 
stellar perturber.  {\bf Type 2 - \toneCtwo\ is the perturber:} This can be denoted as 
[\toneCone, BN] \toneCtwo.  {\bf Type 3 - \toneCone\ is the perturber:} This can 
be denoted as [\toneCtwo, BN] \toneCone.  

Out of all outcomes of our numerical 
experiments we focus on the ones that could
lead to ejection of BN. {\bf Case 1 - BN fly-by:} This is a Type 1 interaction and the outcome 
of interest is ``preservation" where the initial binary members remain unchanged and BN 
flies by after interaction with the initial binary.  
We write this interaction as
\caseone. The arrow
points from the initial to the final configuration. {\bf Case 2 -
  ejection of BN from a binary via exchange with $\bf \theta^1{\rm
    C}_2$:} \casetwo. {\bf Case 3 - ejection of BN from a binary via exchange with
     $\bf \theta^1{\rm C}_1$:} \casethree.  The ``Types" and ``Cases" are different 
     in the fact that for the Types we only take into account the initial conditions 
     and allow all outcomes, whereas, the Cases are more restrictive and only 
     considers the outcomes where BN is ejected leaving behind a bound binary.  

There are 7 parameters that describe the initial conditions of each
binary-single star interaction and since the 3-body problem is chaotic
we must sample over the expected distributions of these parameters:
(1) Eccentricity of the initial binary, $e_i$. For each case we
investigate two extreme distributions: (A) Circular, $e_i=0$, for all
systems, which may be expected if the binaries have recently formed
from a gas disk that has damped out noncircular motions; (B)
Thermal \citep{2003gmbp.book.....H}, $dF_b/de_i= 2 e_i$, where $F_b$ is
the fraction of the binary population. This is an extreme scenario
that would result if binaries have had time to thermalize via stellar
interactions with other cluster stars. The actual situation for ONC
binaries should be between these limits. (2) Semi-major axis of the
initial binary, $a_i$. We assume a flat
distribution \citep{1989ApJ...347..998E} $dF_b/d {\rm log} a_i = 0.208$
from $0.1\,\au$ (approximately the limit resulting from physical
contact) to $6300\,\au$ \citep[the hard-soft
boundary beyond which binaries are expected
to be disrupted by interactions with other cluster stars;][]{2003gmbp.book.....H}.
(3) Initial impact parameter, $b_i$. For each Case and each sampling of $a_i$ we
investigate the full range of impact parameters that can lead to
scattering events strong enough to eject BN with its observed high
velocity. This is achieved by increasing $b_i$ from small values until
the regime where all interactions are weak fly-bys incapable of increasing BN's 
velocity to the observed large value. (4) Initial
relative velocity at infinity, $v_i$, in the frame of the center of
mass of the binary. We assume that the stars have velocities 
following a Maxwellian distribution with a dispersion of 
$\sigma_{\rm{3D}}=3\:\rm{km\:s^{-1}}$ \citep{2008ApJ...676.1109F}. 
We have repeated the numerical experiments with $\sigma_{\rm{3D}}=2\:{\rm \rm
  km\:s^{-1}}$, finding qualitatively similar results. (5) The initial
angle of the orbital angular momentum vector ($\vec L_i$) of the binary with
respect to the velocity vector ($\vec v_i$) of the approaching single star, which
is assumed to be randomly oriented. (6) The initial angle between the
major axis of the binary orbit and the velocity vector of the
approaching single star, which is assumed to be randomly oriented. (7)
The initial orbital phase of the binary, which is assumed to be random.

We find cross-sections ($\Sigma$) of the various outcomes of the Cases 1, 2, \& 3
binary-single interactions numerically using the \code{Fewbody}
software \citep{2004MNRAS.352....1F}, which uses an order $8$
Runge-Kutta integrator, by performing $\sim 10^7$ numerical scattering
experiments to sample the $7$ dimensional parameter space that is needed
to describe the possible interactions.  This large number of numerical
scattering experiments gives us rigorous sampling of all properties in
the dynamical scattering problem, including the initial semimajor axis
of the binary ($a_i$), initial eccentricity ($e_i$), binary orbital
phases, initial velocity at infinity ($v_i$) of the single star, the
initial impact parameter of the encounter ($b_i$), the angle between
the initial major axis relative to $\vec v_i$ and the initial
orientation of the binary ($\vec L_i . \vec v_i$). For
example, we sample $b_i$ as fine as $10^{-4}b_0$, where $b_0$ is the
impact parameter at infinity that results in a closest approach within
$2a_i$. Starting from a small value, $b_i$ is sampled with the above-mentioned 
resolution up to at least $b_{i,{\rm max}} = b_0$. Within this interval smaller intervals 
of $b_i$, $\delta b_i = 10^{-4} b_0$ are chosen. The impact parameter is chosen 
from each of these intervals uniform in the area of the annulus between $b^{'}_{i}$ and 
$b^{'}_{i} + \delta b_i$. 
If a particular final outcome of interest or ``event'' is achieved (in particular \bnvel\ or \truebn\ events, defined below), then contribution of that 
event to the total $\Sigma$ is simply $\delta\Sigma = 2\pi b^{'}_{i} \delta b_i$ 
(\citealt{1996ApJ...467..348M}; but see more recently \citealt{2006ApJ...640.1086F}). 
Afterwards, to ascertain that all 
energetic encounters are sampled, we increase the maximum $b_i$ geometrically 
until $b_{i,{\rm max}} = 100 b_0$. For our case, this large value of $b_{i,{\rm max}}$ corresponds 
to as large a physical distance as the cluster size making sure that all possible 
energetic encounters are captured in the determination of the cross-sections $\Sigma$. 

Using the large ensemble of numerical gravitational scattering experiments 
we evaluate the $\Sigma$s for
outcomes where BN is ejected leaving behind \toneCone\ and \toneCtwo\ 
in a bound binary (henceforth, ``\bnejection" events).  
A subset of the \bnejection\ events where BN is ejected with the observed velocity of
$29\pm3\,\kms$ are called ``\bnvel'' events. We do not put any 
constraints on the binary properties that is left behind for the \bnvel\ events. 
We further calculate the $\Sigma$s of a subset of
\bnvel\ events where the final binary is left with orbital properties
similar to those observed of \toneC, namely, $a =
18.13\pm1.28\,\rm{AU}$, and $e = 0.592\pm0.07$ 
\citep[][henceforth, ``\truebn'' events]{2009A&A...497..195K}. 
The \bnejection\ events are used to explore the velocity distribution of BN if it is ejected via 
a strong binary-single interaction.  
\bnvel\ events, a subset of the \bnejection\ events, show us all possible interactions over a range of 
$a_i$ where BN could have an energy compatible with the observed 
energy. A further subset, the \truebn\ events, give us stronger constraints and indicates the 
range of initial binary properties most likely to create the observed \toneC\ binary 
as well as the runaway BN star.  

\section{Results}
\label{sec:results}
%
%
\begin{figure*}
\begin{center}
\plotone{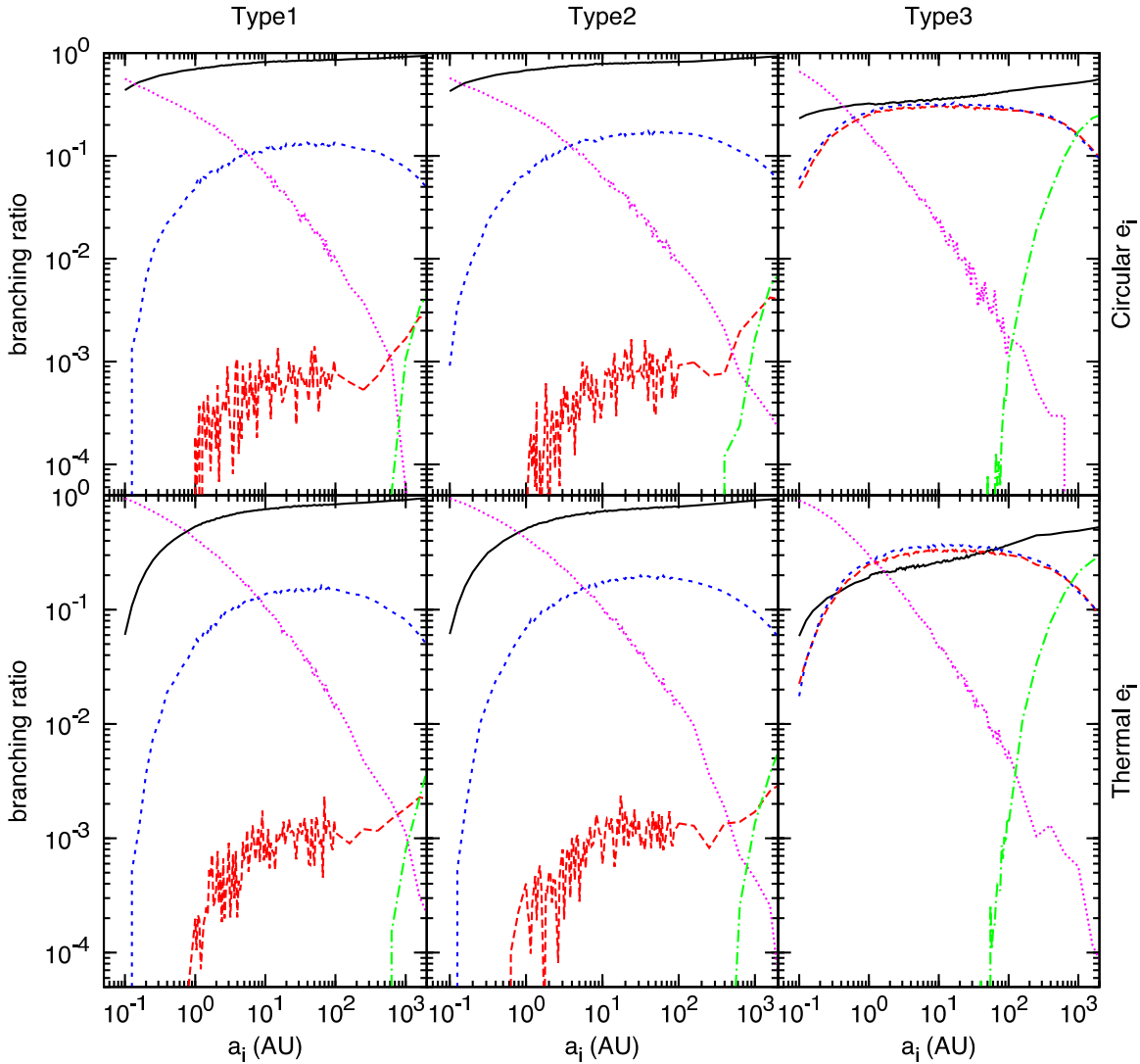}
\caption[branching ratio]{\footnotesize Branching ratios for various outcomes 
in our simulations as a function of $a_i$.  Solid (black), dotted (magenta), 
short-dashed (blue), long-dashed (red), and dash-dotted (green) lines denote 
branching ratios for preservation, collisional outcome, exchange of the perturber 
with the secondary of the initial binary, exchange of the perturber with the initial 
primary, and disruption of the binary, respectively.  The top and bottom panels are 
for Circular and Thermal $e_i$ distributions, respectively.  In both top and bottom 
panels three panels from left to right denote Types 1, 2, and 3, respectively.  
}
\label{fig:branching}
\end{center}
\end{figure*}
%
In this section we present the key results of our numerical experiments.  We start with 
overall outcomes of all our numerical experiments for all Cases and $e_i$ distributions 
and then increasingly focus our attention towards the observed BN-\toneC\ system and 
compare its various properties with those predicted from our simulations.  

Figure\ \ref{fig:branching} shows the branching ratios of all outcomes in general from our numerical 
experiments.  A handful of interesting aspects are evident in the branching ratios for the given 
masses of the 3 stars involved in these interactions.  

Disruption (or ionization) of the initial 
binary happens only when the binary is dynamically soft \citep{1983ApJ...268..319H}, i.e., the 
value of the binary binding energy is 
$\lesssim$ the kinetic energy of the perturber.  This is achieved at large $a_i \gtrsim 300\,\au$, 
for Types 1 and 2.  For Type 3 ionization can happen at relatively smaller 
$a_i \sim 50\,\au$ due to the higher mass of the perturber and relatively lower binding 
energy of the initial binary.  Nevertheless, even for Type 3, branching ratio for 
ionization becomes comparable or greater than exchange outcomes only at $a_i \sim 10^3\,\au$.  

For Types 1 and 2, exchange with the primary is very unlikely, since here the primary is 
significantly more massive ($38.2\,\msun$) than the secondary ($8.8$ and $8.2\,\msun$ for interactions 
of Types 1 and 2, respectively).  In interaction of Type 3, since the initial binary consists of two stars with comparable 
masses, both exchanges are almost equally likely.  For Type 3 the fraction of exchange outcomes 
is comparable to the fraction of fly-by events for a large range of $a_i$ taking into 
account sufficiently strong encounters (see \S\ref{sec:method} for the value of the maximum impact 
parameter).  The fraction of fly-by outcomes is of course formally 
infinite since one can always use a sufficiently large impact parameter where nothing but a weak fly-by 
is the outcome.   

Collisional outcomes are comparable 
with exchanges only for sufficiently small $a_i$ values.  In interactions of Type 1, preservation is the channel 
that can produce the observed BN-\toneC\ system.  For Type 1 for both $e_i$ distributions 
collisions become important for $a_i \lesssim 1\,\au$.  For interactions of Types 2 and 3, exchange of 
the perturber with the primary is the channel that can produce the observed BN-\toneC\ system.  
For Type 2, collisions become comparable with the BN-\toneC\ producing channel 
for $a_i \lesssim$ a few AU.  Whereas, for interactions of Type 3, collisions remain comparable to the 
BN-\toneC\ producing channel for $a_i \lesssim 1\,\au$.  In interactions of all Types collisions happen more 
often for the Thermal $e_i$ distribution since for the Thermal $e_i$ distribution the pericenter 
distances for the stars in binary can be much smaller than that for the Circular $e_i$ distribution 
for any given $a_i$.  We later show (\S\ref{sec:cross-sections}) that \truebn\ events happen in 
a given range of $a_i$ for a given interaction Type.  For all interaction Types and $e_i$ distributions 
collisions have much lower branching ratios compared to the branching ratios for the BN-\toneC\ 
production channels for the ranges of $a_i$ where \truebn\ events can occur.           

In the following sections we increasingly focus on outcomes that are similar to the observed 
BN-\toneC\ system.  We first present 
results for all events where the BN star is ejected (\bnejection, \S\ref{sec:vdist}).  Then we present results for all 
outcomes where the BN star is ejected with a velocity within the observed range of $29 \pm 3\,\kms$ 
(\bnvel, \S\ref{sec:eratio}).  We then restrict our attention to only a subset of the \bnvel\ events where the 
final binary has properties similar to the observed \toneC\ binary (\truebn, \S\ref{sec:cross-sections} and \S\ref{sec:angle}).

%
%
%
%
\begin{figure*}
\begin{center}
\plotone{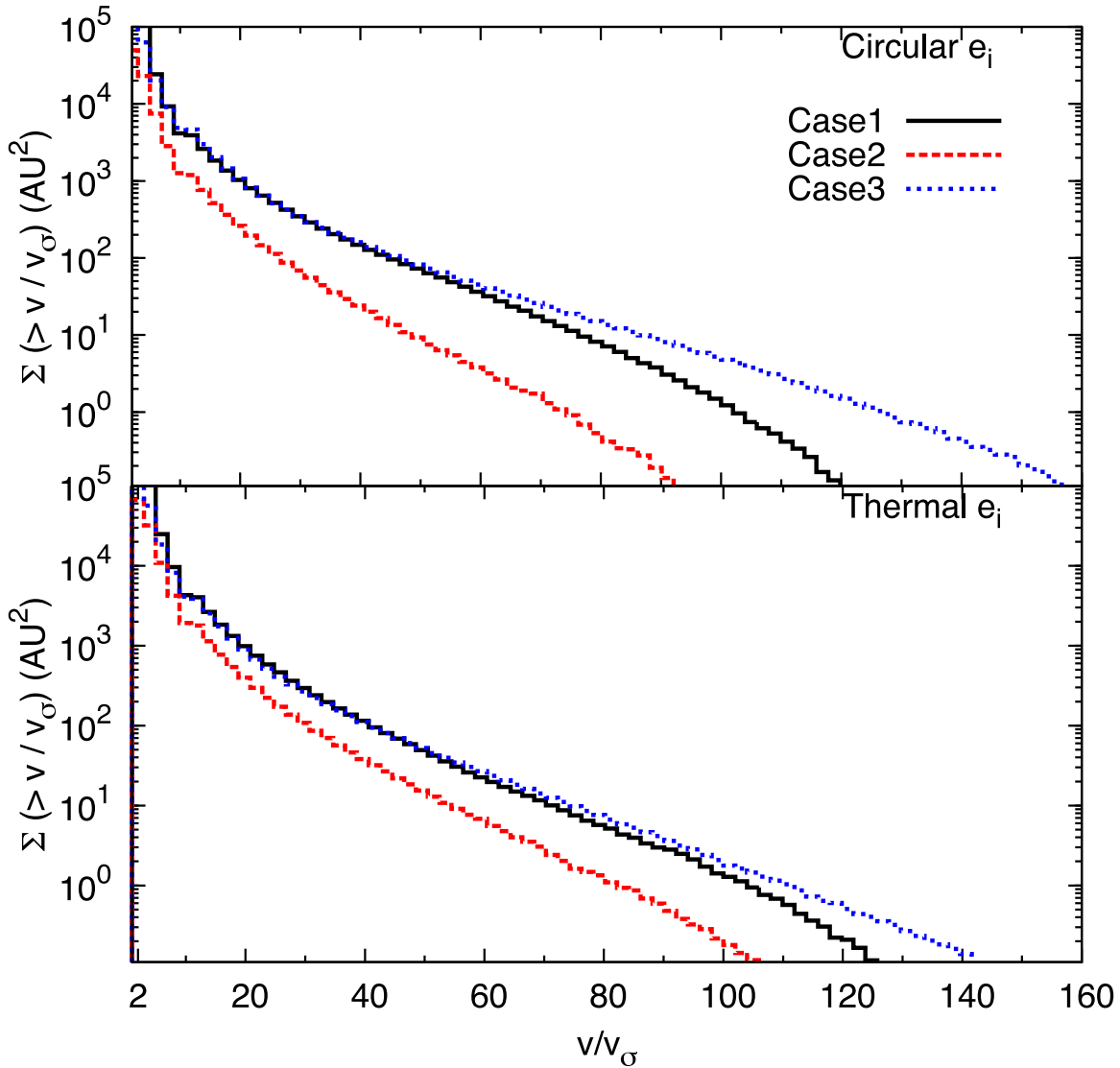} 
\caption[$v_{BN}$ distribution]{\footnotesize Cumulative distribution 
of $\Sigma$ as a function of the velocity of the ejected star for all cases where 
BN is ejected. The velocities are given in units of the velocity dispersion 
($v_\sigma = 3\,\kms$).  
Top panel is for the Circular $e_i$ distribution. The bottom panel is the 
same but for the Thermal $e_i$ distribution. 
The solid (black), dashed (red), and dotted (blue) lines in both panels denote 
Cases 1, 2, and 3, respectively. For each case the cross-section is 
calculated using 
$\int^{v}_{v^{'}=2v_{\sigma}} \int_{a_i} P(a_i) \frac{d\Sigma(a_i, E_{\rm{ratio}}, e)}{dv^{'}} da_i dv^{'}$ for all 
events where BN is ejected. For all cases $P(a_i)da_i =
\delta {\rm log}(a_i)/{\rm log}(6310/0.1)$ is used (see text). We use a cut-off for the ejection 
velocity at $v = 2v_\sigma$ since for $v\sim v_\sigma$ 
$\Sigma(>v/v_\sigma)$ becomes very large simply due to distant fly-by interactions 
in Case 1. Note that for all cases dynamical interactions can 
increase the velocity of the ejected star by large factors relative to the velocity dispersion.   
}
\label{fig:vdist}
\end{center}
\end{figure*}
%

\subsection{Velocity Distribution of the Ejected BN Star}
\label{sec:vdist}
Increasing the kinetic energy of BN by about two orders of
magnitude compared to the value expected given the
ONC's velocity dispersion is at the heart of the problem. 
Hence, we focus on the velocity distribution of BN following ejection.  
In addition, we focus on energy considerations for the scattering problem, 
especially the ratio of the kinetic energy of
BN's ejection to the total energy of the binary left behind.  

First we explore given the masses of the three stars in the interaction, and given that BN 
is ejected leaving the other stars in a binary, how likely it is for BN to acquire a velocity significantly higher than the 
velocity dispersion ($v_\sigma = 3\,\kms$) in the ONC.  
We calculate the cross-section $\Sigma_{\bnejection}$ for \bnejection\ events for a given $a_i$, 
for each Cases 1--3, and each $e_i$ distribution.  The overall cross-section for \bnejection\ 
events for {\it any} $a_i$ is calculated by multiplying $\Sigma_{\bnejection}$ with the probability of finding 
an initial binary with that $a_i$ assuming the semimajor axis distribution for
binaries is flat in logarithmic intervals within the physical limits
discussed in \S\ref{sec:method}. Thus, the normalized
$\Sigma_{\bnejection}$ is 
$\int P(a_i) \Sigma_{\bnejection} (a_i, v_{\rm{BN}}, e) d a_i$, 
where, 
$P(a_i)da_i = d {\rm log} (a_i) / {\rm log} (6310/0.1)$. 

Figure\ \ref{fig:vdist} shows the cumulative distribution of $\Sigma_{\bnejection} (v_{\rm{BN}})$ as a function 
of BN's velocity ($v_{\rm{BN}}$) calculated using $\int^{v}_{v_{\rm{BN}} = 2v_\sigma} dv_{\rm{BN}} \times d\Sigma_{\bnejection}/dv_{\rm{BN}}$.  We find that binary-single interactions involving the three stars in question 
can eject the BN star with velocities that can exceed $v_\sigma$ by more than two orders of 
magnitude.  However, the cross-sections for such events reduce as $v_{\rm{BN}}$ increases.  
For both $e_i$ distributions Case 3 shows a higher fraction of high-velocity ejection 
events.  This is because in Case 3 the perturber is the most massive star in the triplet, 
the one that finally becomes the \toneCone\ star.  Hence, the total available energy is higher 
in Case 3 events.     

%
%
\begin{figure*}
\begin{center}
\plotone{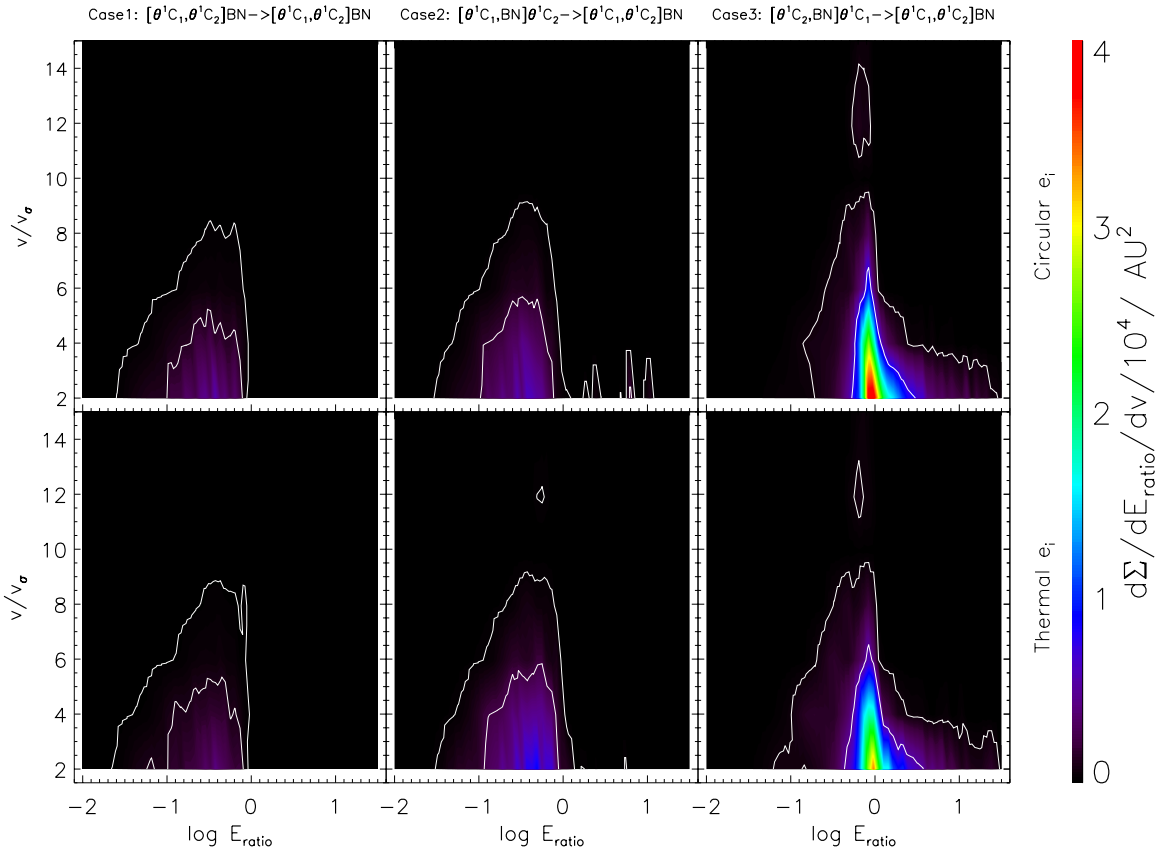}
\caption[$E_{\rm{ratio}} vs v$]{\footnotesize 2D distribution of $\Sigma$ 
as a function of energy ratio
$E_{\rm{ratio}}\equiv T_{\rm ejection}/|E_{\rm binary}|$ and 
velocity ($v$) of the runaway star BN for all events where BN is ejected. 
The top panels are for the Circular
$e_i$ distributions for Cases 1--3 (left to right, respectively).  The
bottom panels are for the same, but for the Thermal $e_i$
distribution.  The colors denote $d \Sigma / dE_{\rm{ratio}} / dv$.
For each bin the total cross-section is calculated using $\int P(a_i)
\Sigma(a_i, E_{\rm{ratio}}, e) da_i$ for the events where BN is ejected 
satisfying the ranges of $E_{\rm{ratio}}$ and $v$ in that bin.  $P(a_i)da_i =
\delta {\rm log}(a_i)/{\rm log}(6310/0.1)$ is used (see text). 
For both $e_i$ distributions Case3 shows higher values of $E_{\rm{ratio}}$ than 
Cases 1 and 2. This is because in Case 3 the perturber is a more massive 
star (the one that will finally become \toneCone~). For all Cases the highest 
velocity increase for the runaway star is between $E_{\rm{ratio}} = 0.1$ and $1$.
}
\label{fig:eratio_v}
\end{center}
\end{figure*}
%
The problem of binary-single scattering can be understood by comparing 
the kinetic energy and the potential energy of the systems since the 
outcomes differ qualitatively depending on the relative values of these 
quantities \citep[e.g.,][]{1983ApJ...268..319H}.  
We explore if BN is ejected with $v_{\rm{BN}} \geq 2v_\sigma$, then what is the distribution 
of cross-section for the various Cases and $e_i$-distributions as a function of the $E_{\rm{ratio}}$ and 
the velocity of ejection, $v_{\rm{BN}}$.  Here, 
$E_{\rm{ratio}}\equiv T_{\rm ejection}/|E_{\rm binary}|$ is the ratio of the final kinetic energy 
($T_{\rm ejection}$) of {\it both} the single star and the binary star system (based on the motion 
of its center of mass) to the total energy (gravitational energy plus kinetic energy of orbital motion) of the binary 
($E_{\rm{binary}}$). Figure\ \ref{fig:eratio_v} shows a $2D$ distribution of the overall 
$\Sigma_{\bnejection}$ for any $a_i$ as a function of $v_{\rm{BN}}$ and $E_{\rm{ratio}}$ 
for all Cases 1--3, and all $e_i$ distributions.  The highest ejection velocities for 
the BN star happens for $E_{\rm{ratio}}$ between 0.1 and 1 for all Cases. 
Note that the distributions for Cases 1 and 2 are very similar. This is due to the 
similarity in masses of BN and \toneCtwo. However, in Case 3 
the most massive star is the perturber.  In addition, the most massive star exchanges into 
the final binary increasing the binding energy of the final binary star system.  Hence, 
there is a tail for high $E_{\rm{ratio}}$ values where BN can still be ejected with a high velocity. 
Even for the more energetic Case 3 events, the maximum velocity for ejections 
are achieved within a narrow range of $0.1 < E_{\rm{ratio}} < 1$.    

A high value ($\sim 10$) of the ratio between the velocities of the incoming single 
star (expected to be near $v_\sigma$) and the ejected single star is needed 
to create runaway stars by definition.  This is possible for strong encounters 
involving a binary and a single star where a fraction of binding energy of the 
binary is converted into the kinetic energy ($T$) of the ejected star. 
If the final outcome is again a binary and the runaway ejected star (the binary 
members are not required to remain the initial ones, 
e.g., for Cases 2 and 3) 
then the kinetic energy of the stars undergoing dynamical ejection,
$T_{\rm ejection}= (1/2) m_{\rm BN} v_{\rm BN}^2 + (1/2) (m_{\rm
  \theta^1C_1}+ m_{\rm \theta^1C_2}) v_{\rm \theta^1C}^2\rightarrow
(6.9\pm2.7)\times 10^{46}\:{\rm erg} + (1.2\pm0.5)\times 10^{46}\:{\rm
  erg} = (8.1\pm2.8)\times 10^{46}\:{\rm erg}$ (here $\rightarrow$
indicates the observed values) is expected to be less than the magnitude of the
total energy of the resulting binary, $|E_{\rm binary}|= G m_{\rm
  \theta^1C_1} m_{\rm \theta^1C_2}/(2a)\rightarrow (16.4\pm4.9)\times
10^{46}\:{\rm erg}$. In addition to requiring $E_{\rm ratio}<1$, one
also expects it to achieve a value of order unity, i.e. not too much
less than one. 
For $E_{\rm {ratio}}\ll 1$, collisional outcomes 
  dominate \citep[Figure\ \ref{fig:branching}; also see e.g.,][]{1983ApJ...268..319H,2004MNRAS.352....1F}. 
  On the other hand, for $E_{\rm {ratio}}\gg 1$ 
outcomes with a disruption of the 
  binary dominates creating 3 single stars (Figure\ \ref{fig:branching}).  
  The observed value for the BN-\toneC\ system is $E_{\rm
  ratio} \rightarrow 0.49\pm 0.22$, consistent with it being a result of a binary-single 
  interaction.      
  
%
%
%
\begin{figure*}
\begin{center}
\plotone{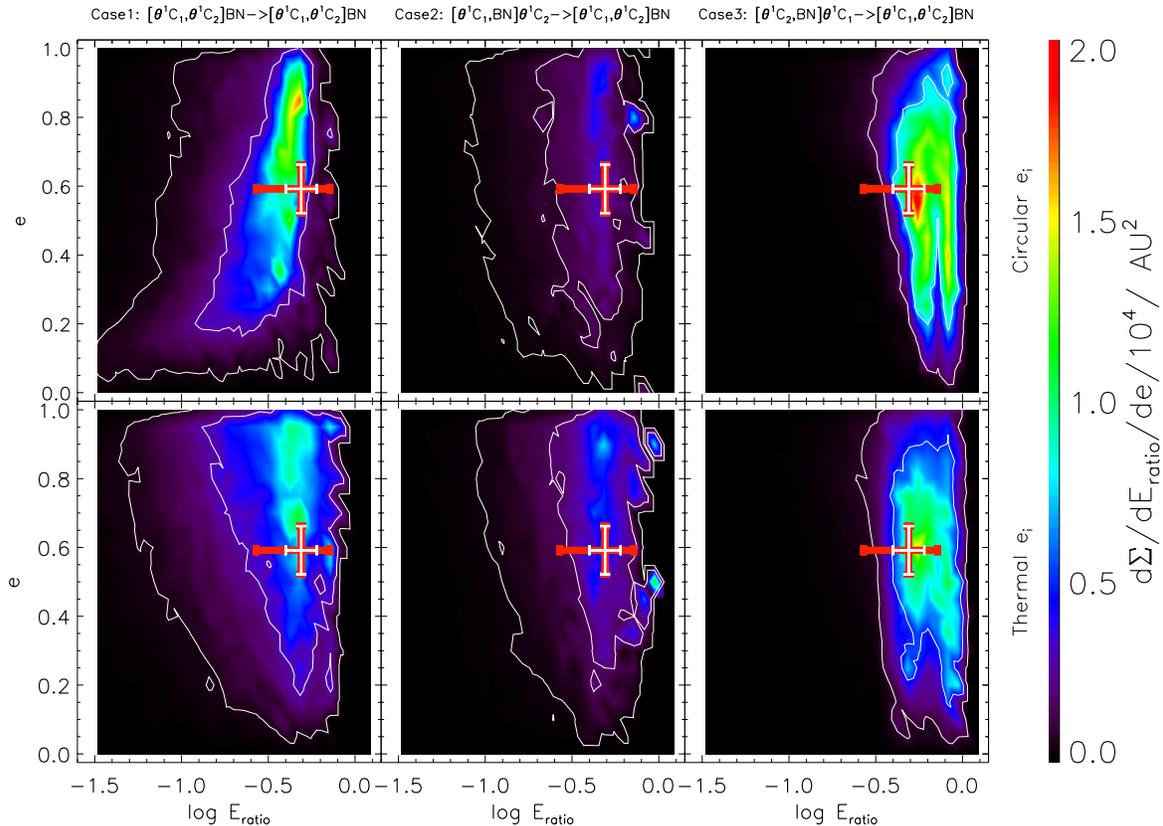}
\caption[$E_{\rm{ratio}} vs e$]{\footnotesize 2D distribution of $\Sigma$ as a function of energy ratio
$E_{\rm{ratio}}\equiv T_{\rm ejection}/|E_{\rm binary}|$ and final
eccentricity of the resulting binary $e$ for the \bnvel~events for all
cases and $e_i$ distributions.  The top panels are for the Circular
$e_i$ distributions for Cases 1--3 (left to right, respectively).  The
bottom panels are for the same, but for the Thermal $e_i$
distribution.  The colors denote $d \Sigma / dE_{\rm{ratio}} / de$.
For each bin the total cross-section is calculated using $\int P(a_i)
\Sigma(a_i, E_{\rm{ratio}}, e) da_i$ for the \bnvel~events satisfying
the ranges of $E_{\rm{ratio}}$ and $e$ in that bin.  $P(a_i)da_i =
\delta {\rm log}(a_i)/{\rm log}(6310/0.1)$ is used (see text). The point and
errorbars show the observed BN-\toneC\ system. The larger horizontal
errorbars (red) denote the $E_{\rm{ratio}}$ errors including
contribution from the mass errors of the stars, whereas the shorter
(white) errorbars denote the same if the central mass values are chosen
and no contribution from mass errors are included (consistent with the
numerical experiments).  For all cases and all $e_i$-distributions the
observed system properties lie within the $1\sigma$ contours of the
$\Sigma$-distribution, with Case 3 Circular $e_i$ being somewhat more favored (see text).}
\label{fig:eratio_e}
\end{center}
\end{figure*}
%

\subsection{Kinetic Energy of Ejection and Eccentricity of Binary}
\label{sec:eratio}
We now focus on the subset of \bnejection\ events (\bnvel) where BN is ejected with the observed
velocity of $29\pm3$ (\S\ref{sec:method}) via any one of the
Cases 1--3 (a small strip in the vertical axis in Figure\ \ref{fig:eratio_v}). 
The \toneC\ orbit is eccentric ($e \approx 0.6$). 
Indeed, strong encounters are expected to leave behind binaries with 
generally high eccentricities. We now want to see the $2D$ distribution of 
overall cross-section for all \bnvel\ events for any $a_i$ as a function of 
$E_{\rm{ratio}}$ and the final eccentricity.    

In Figure \ref{fig:eratio_e} we plot the 2D distributions of $\Sigma$ for the
\bnvel\ events in the $E_{\rm ratio}$ versus final $e$ plane for Cases
1, 2, \& 3 for both the Circular and Thermal $e_i$ distributions. The
observed values of the BN-\toneC\ system are also shown. These overlap
within the $1\sigma$ contours for all cases. The energy of the
\toneC\ binary is just what we would expect if it had ejected BN at
the observed velocity. Its high eccentricity, $\approx 0.6$, is also
naturally explained by the recent ejection of BN 
since during ejection the potential
of the system is changing rapidly.

Note that if the \toneC\ binary was unrelated to BN, then
$E_{\rm{ratio}}$ could have values in a large range spanning many
orders of magnitude.
For example, the range in $a$ for the binary is determined by contact
($\sim$ a few stellar radii) to the hard-soft boundary in ONC ($\sim
6000\,\au$ for the velocity dispersion in ONC). Thus $E_{\rm{binary}}$
for \toneC\, if unrelated to BN's ejection, could be expected to have
values anywhere between $\sim 10^{44}$ and $10^{49}\,\rm{ergs}$.  It
is thus very interesting to find the observed BN-\toneC\ system with
energies so close to the expected energies if they have had a
binary-single scattering encounter in the past. The likelihood of this
occuring simply by chance rather than by being caused by interaction
with BN is explored in \S\ref{sec:probability}.

%
%

\begin{figure*}
\begin{center}
\plotone{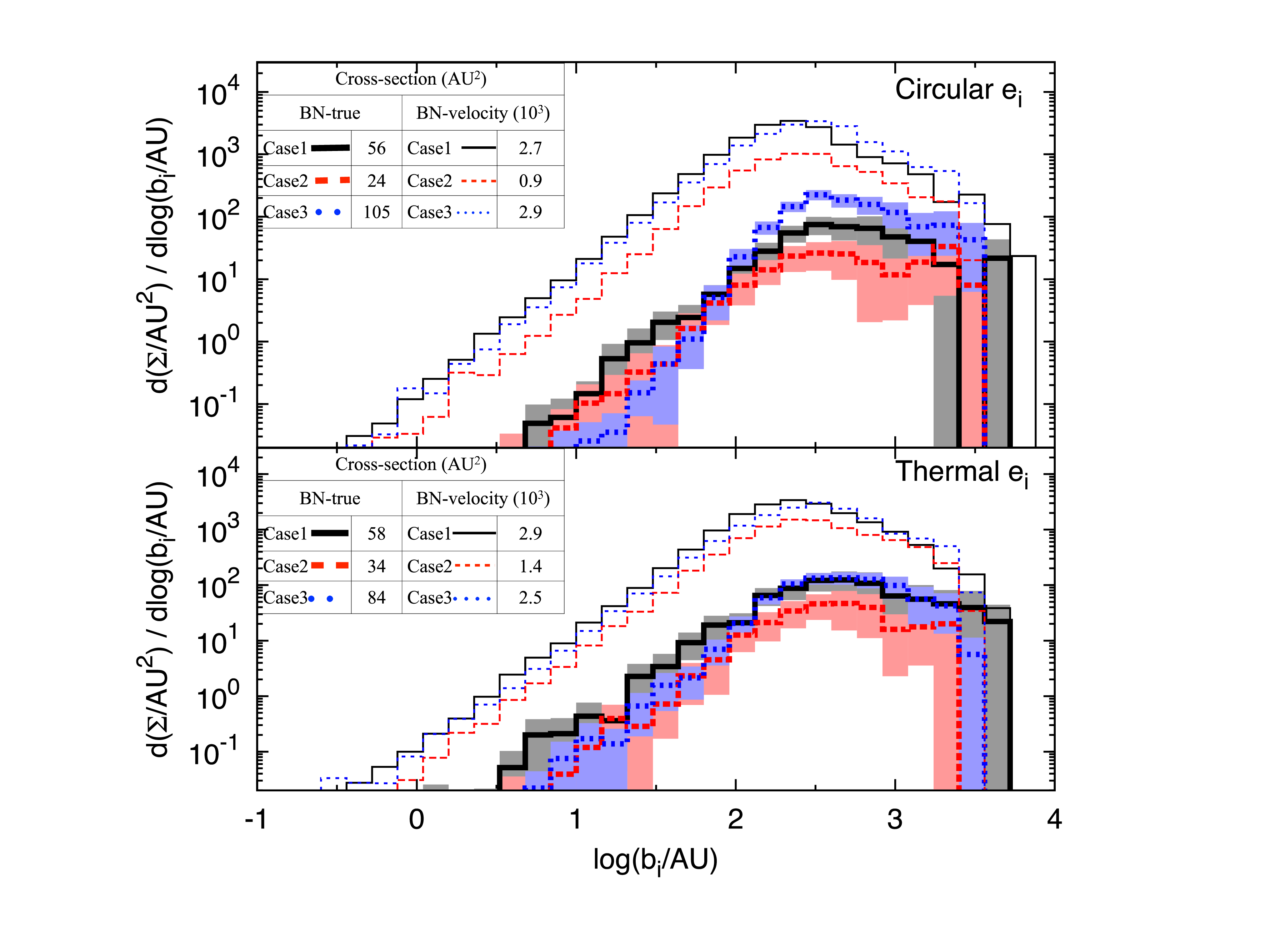}
\caption[$b_i$ vs $\Sigma$]{\footnotesize Cross-section, $\Sigma$, vs initial impact parameter, $b_i$, of
interaction.  The top and bottom panels show results for the Circular
and Thermal $e_i$ distributions, respectively.  Thin and thick lines
represent \bnvel\ and \truebn\ events, respectively.  Black (solid),
red (long-dashed), and blue (short-dashed) lines denote Cases 1, 2,
and 3, respectively. The power-law increase in cross-section for low
$b_i$ is simply due to geometry. Most interactions take place within
$b_i\approx300\,\au$. However, interactions with $b_i$ much larger
($\gtrsim 10^3\,\au$) may still contribute.  The color-matched shaded
regions around each \truebn\ histogram denote Poisson errors in estimating
$\Sigma$.  The inset tables denote the total $\Sigma$ values for \bnvel\ and
\truebn\ events for all cases and $e_i$-distributions.
All three cases can contribute towards producing the observed
BN-\toneC\ system, however, Case 3 is preferred by a factor of $\sim 2$. }
\label{fig:b_sigma}
\end{center}
\end{figure*}


\begin{figure*}
\begin{center}
\plotone{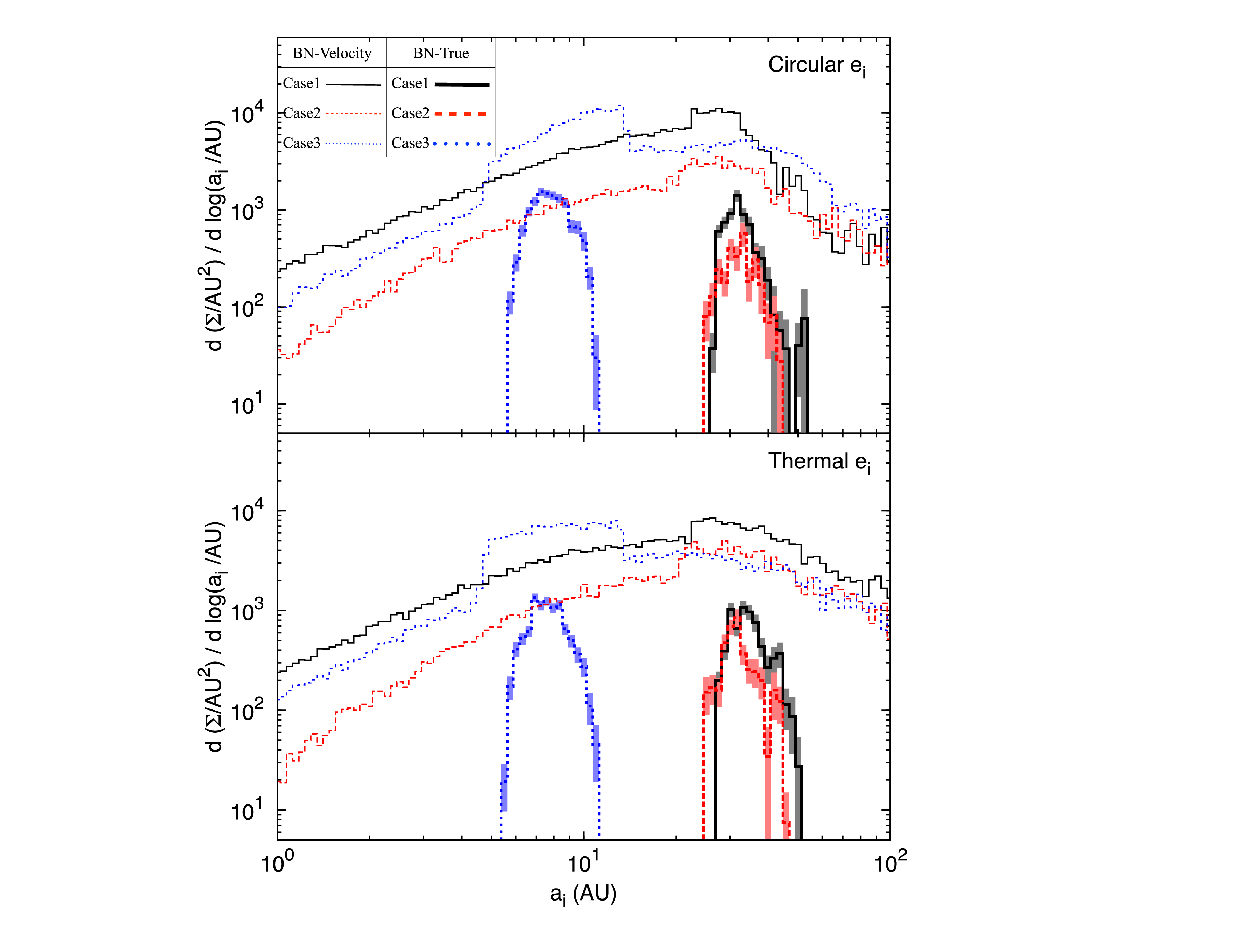}
\caption[$a_i$ vs $\Sigma$]{\footnotesize Same as Figure~3, but for $\Sigma$ as a function of the initial
semimajor axis, $a_i$, of the binary.  We find that all three cases
and $e_i$-distributions can contribute to the production of the
observed system over different $a_i$-ranges, although Case 3 has a
relatively higher integrated $\Sigma$ for both $e_i$-distributions and selects from a quite distinct range of $a_i\simeq 8\pm 2$~AU.}
\label{fig:a_sigma}
\end{center}
\end{figure*}

\subsection{Cross-Sections of \bnvel\ and \truebn\ Events:}
\label{sec:cross-sections}
We evaluate the cross-sections for outcomes where BN is ejected with
the observed velocity of $29\pm3\,\kms$ (``\bnvel'' events). We
further calculate the cross-section of a subset of \bnvel\ events
where the final binary is left with orbital properties similar to
those observed of \toneC, namely, $a = 18.13\pm1.28\,\rm{AU}$, and $e
= 0.592\pm0.07$ (\citealt{2009A&A...497..195K}; ``\truebn'' events).

For Cases 1, 2, 3 with Circular initial $e_i$ distribution, 
$\Sigma_{\rm BN-Velocity} = (2.7, 0.94, 2.9)\times 10^3\:{\rm AU}^2$, 
while for Thermal initial $e_i$ distribution, 
$\Sigma_{\rm BN-Velocity} = (2.9, 1.4, 2.5)\times 10^3\:{\rm AU}^2$. For Cases
1, 2, 3 with Circular $e_i$-distribution, $\Sigma_{\rm BN-True}
= 56, 24, 105\:{\rm AU}^2$, while for thermal $e_i$
distribution, $\Sigma_{\rm BN-True} = 58, 34, 84\:{\rm AU}^2$. The
three cases have similar cross-sections, with Case 3, i.e. \casethree,
somewhat more preferred.  

The cross-sections for \bnvel\ and \truebn\ events as a function of
$b_i$, are shown in Figure\ \ref{fig:b_sigma}. The cross-sections for a 
given $b_i$ and for all explored $a_i$ are normalized using the probability, 
$P(a_i)$, of finding a binary with semimajor axis $a_i$, 
assuming a semimajor axis distribution flat in log intervals. 
For small values of $b_i\lesssim300\,\au$, the cross-sections grow geometrically as 
$b_i^2$ and then decline. This decline, seen in all cases, is simply due to the fact that the 
interactions are happening at larger and larger impact parameters and at some 
point no interactions are expected to be strong enough to increase the energy of 
the ejected star to the observed value of BN. The areas under the histograms are 
the total cross-sections for the \bnvel\ and \truebn\ events.  
Figure\ \ref{fig:b_sigma} also includes a table summarizing the total
cross-sections.

The cross-sections for \bnvel\ and \truebn\ events as a function of
$a_i$ are shown in Figure~\ref{fig:a_sigma}.
We find that \bnvel\ events happen via binary-single scattering
encounters for a large range of $a_i$, but \truebn\ events place much
tighter constraints on $a_i$. For example, Case 3, which is the most
favored, requires the initial binary [$\theta^1{\rm C}_2$, BN],
i.e. two approximately equal-mass stars of $\sim 9\,\msun$, to have
originally had $a_i\simeq 8\pm 2$~AU.  Note that depending on the case, 
different ranges of $a_i$ contribute towards creating systems similar 
to the observed BN-\toneC\ system.  Moreover, note that the $a_i$ ranges 
where the \truebn\ events occur via the three Cases 1 -- 3, ejection of 
BN is the most likely outcome among all other possible outcomes 
(except of course weak fly-bys; Figure\ \ref{fig:branching}).  This gives us 
further confidence in the scenario that the BN-\toneC\ system has been 
created via a strong binary-single interaction involving these three stars.     

Throughout this study we have used the central values of the estimated
masses for the three stars.  Dynamically there should be no
qualitative difference in the outcomes if the masses are changed
within the mass errors.  However, note that the masses of
$\theta^1{\rm C}_2$ and BN are comparable.  In fact the error ranges
actually overlap.  Due to the comparable masses, dynamically there is
only a small difference between ejection of BN and ejection of
$\theta^1{\rm C}_2$, as named here. This is reflected in our results to some 
degree. Cases 1 and 2
contribute towards creating the BN-\toneC\ system over very similar
$a_i$ ranges.  Their contributions are also comparable.  The small
differences in the $a_i$ range and $\Sigma_\truebn$ come from the small
difference between $\theta^1{\rm C}_2$ and BN's assumed masses and also to
some extent the scenario, a fly-by being more likely compared to an
exchange if all else is kept unchanged. In fact, similar results will be 
recovered if BN and \toneCtwo\ are interchanged among themselves. 
However, in that case, the definitions of Cases 1 and 2 will also be interchanged. 

%

\begin{figure*}
\begin{center}
\plotone{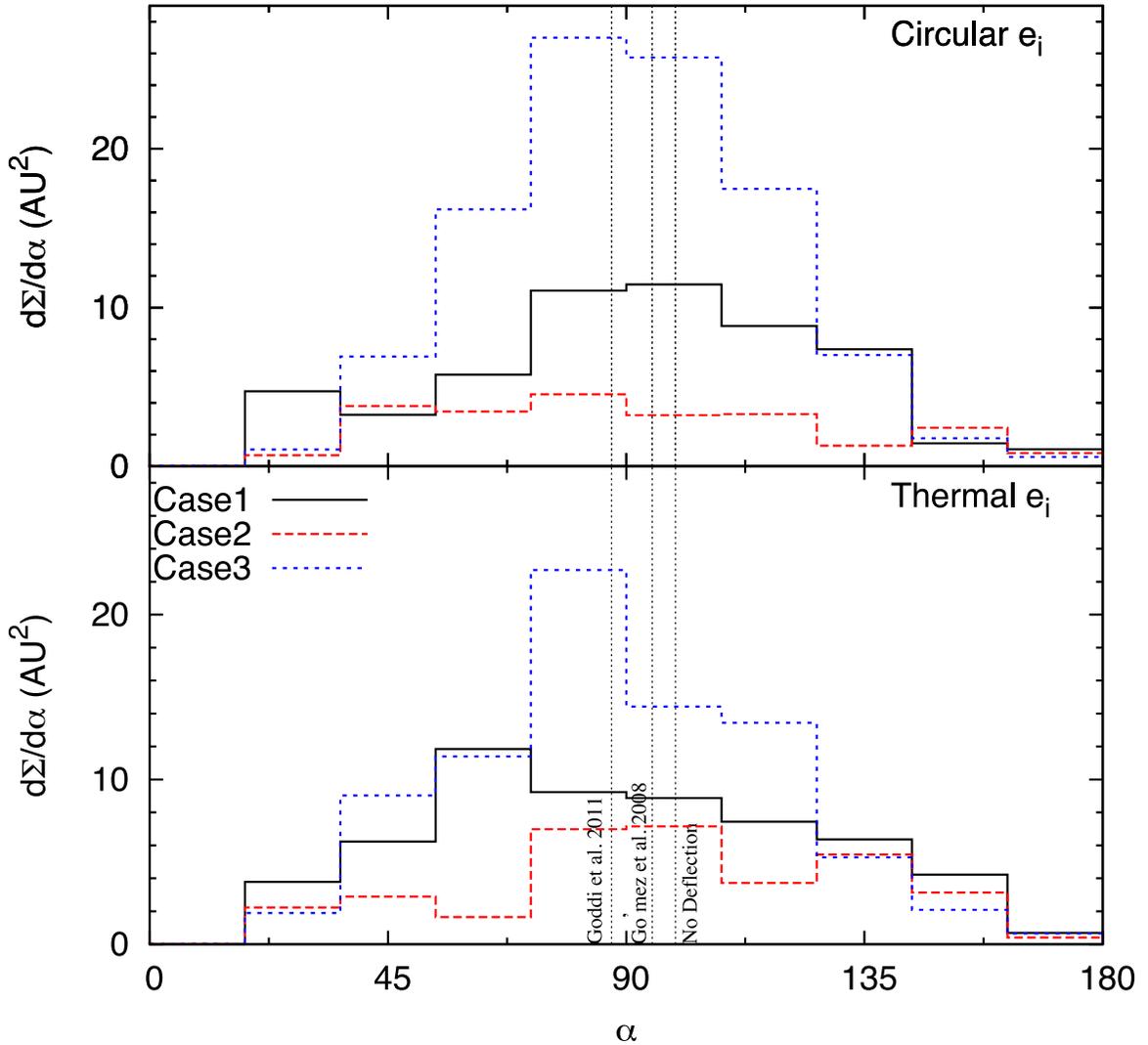}
\caption[$\alpha$ vs $\Sigma$]{\footnotesize Cross-section, $\Sigma$, vs the angle 
($\alpha$) between the $3D$ velocity of the runaway BN star $\vec{v}_{\rm{BN}}$ and 
the angular momentum vector $\vec{L}_{\theta^1{\rm C}}$ of the \toneC~binary. The solid (black), 
dashed (red), and dotted (blue) lines denote Cases 1, 2, and 3, respectively, for \truebn\ events 
(distributions of \bnvel\ events are very similar).
The top and 
bottom panels are for the Circular and Thermal $e_i$ distributions, respectively. The vertical 
dotted (black) lines show three values of $\alpha$ for the observed BN-$\theta^1\rm{C}$ system where 
the values are calculated using $\vec{v}_{\rm{BN}}$ measurements by 
\citet{2011ApJ...728...15G, 2008ApJ...685..333G} and assuming no deflection of 
BN after its ejection from the \toneC~binary. 
Orientation 
of the \toneC~binary is obtained from \citet{2009A&A...497..195K}. 
The predicted distributions from our simulations 
are consistent with the estimated values of $\alpha$ for the observed BN-$\theta^1\rm{C}$ system. 
}
\label{fig:alpha_sigma}
\end{center}
\end{figure*}
\subsection{Orientation of the Orbital Plane of \toneC\ Binary Relative to the Direction of BN's Motion}
\label{sec:angle}
One additional variable in the dynamical ejection problem is the 
angle $\alpha$ between the angular momentum vector, $\vec{L}_{\theta^1{\rm C}}$, 
of the \toneC\ binary and the $3D$ velocity vector of BN, $\vec{v}_{\rm{BN}}$.  
Figure\ \ref{fig:alpha_sigma} shows the distribution of $\Sigma_{\truebn}$ for all \truebn\ 
events for {\it any} $a_i$ as a function of $\alpha$ for all Cases 1--3 and all $e_i$ distributions.  
The distributions of $\alpha$ are quite broad for Cases 1 and 2.  In comparison, for Case 3 there 
is a strong peak near $\alpha=90^\circ$.  

To calculate the observed value of $\alpha$ for comparison with the predictions of our numerical 
results we adopt different measured values in existing literature.  Orientation of $\vec{L}_{\theta^1{\rm C}}$ 
is calculated using data given in \citet{2009A&A...497..195K}.  The LSR velocity of BN is obtained 
from \citet{1983ApJ...275..201S}.  There are two independent proper-motion measurements 
for BN.  Adopting the values given in \citet{2008ApJ...685..333G} we find 
$\alpha = 95^\circ$.  Adopting the values given in \citet{2011ApJ...728...15G} we find a slightly 
different value of $\alpha = 87^\circ$.  If the direction of BN's motion in the sky-plane is obtained by simply 
joining the expected position of the binary-single encounter and BN's current position, then 
$\alpha = 99^\circ$.  
(this may be a more accurate value, since we expect BN to have
suffered a recent change in its proper motion vector via interaction
with source {\it I}, see below)  
Note all of the above values of
$\alpha$ for the observed BN-\toneC\ system are consistent with the
predicted distribution of $\alpha$ from our numerical experiments,
and give some support for the ejection having resulted via Case 3.

%
%
\section{Tests for the Ejection Scenario of BN from \toneC\ and Probability of Chance Agreement}
\label{sec:probability}

The system which ejected BN must be located along BN's past trajectory
and have a total mass $\gtrsim 2 m_{\rm BN}$. These conditions are
potentially satisfied for \toneC, the 3 other Trapezium stars $\rm
\theta^1A$, $\rm \theta^1B$, $\rm \theta^1D$, another ONC member
$\theta^2A$, and probably for source {\it I} (assuming it is the main
source of luminosity in the KL nebula). Indeed, a number of authors
have argued BN was launched from source {\it
  I} \citep{2005AJ....129.2281B, 2008ApJ...685..333G}. However, as we
now discuss, there are 6 independent properties of \toneC\ that have
the values expected if it were the binary left behind after ejecting
BN 
(7 if we assume ejection via Case 3 and include the angle $\alpha$
between the angular momentum vector, $\vec{L}_{\theta^1{\rm C}}$, of
the \toneC\ binary and the $3D$ velocity vector of BN,
$\vec{v}_{\rm{BN}}$). To consider the likelihood that all of these
properties of the BN-\toneC\ system are as observed {\it by chance},
we take that to be our null hypothesis.  Given that BN has the runaway
velocity, for each of these properties we assign a probability that
\toneC\ has its values by chance to finally calculate the composite
probability of chance agreement of \toneC's properties with those
expected from a binary-single scattering scenario. We discuss these
properties and our
estimates of the chance-agreement 
probabilities below. These probabilities are summarized in Table\ \ref{tab:properties} 
along with the values predicted by the binary-single ejection scenario, and the 
observed values of the BN-\toneC\ properties.   

(1) {\it ONC-Frame Proper Motion in Declination ($\mu_{\rm
  \delta,ONC}({\rm \theta^1C})$):} If \toneC\ ejected BN, then, in the
frame of the center of mass of the pre-ejection triple, the predicted value of $\mu_{\rm
  \delta,T}({\rm \theta^1C}) = -(m_{\rm BN}/m_{\rm \theta^1C})
\mu_{\rm \delta,T}({\rm BN}) \rightarrow
-([9.3\pm2.0\,\msun]/[47\pm4\,\msun]) 11.7\pm1.3\:{\rm mas\:yr^{-1}}
\rightarrow -2.3\pm0.6\:{\rm mas\:yr^{-1}}$. Here we have used the
luminosity-based mass estimate for BN \citep{2004ApJ...607L..47T} and
the proper motion measurements of \citet{2008ApJ...685..333G} for BN,
including a 0.70~mas/yr uncertainty of the motion in declination of
the pre-ejection triple with respect to the ONC frame.  The predicted
value of the ONC-frame motion of \toneC\ is then 
$\mu_{\rm \delta,ONC}({\rm \theta^1C})[{\rm predicted}] = -2.3 \pm 0.9\:{\rm mas\:yr^{-1}}$, 
with the error increasing again because of the uncertain motion of the pre-ejection 
triple\footnote{Note that here and for the other \toneC\ properties we have adopted 
$1\sigma$ errors when possible, but not all physical properties have well-defined 
uncertainties: e.g., the model dependent $m_{\rm BN}$ estimate given its observed luminosity.}.
The observed value \citep{1988AJ.....95.1744V} is $\mu_{\rm
  \delta,ONC}(\theta^1C)[{\rm observed}] = -1.8\pm 0.2 \:{\rm mas\:yr^{-1}}$. Given the
observed \citep{1988AJ.....95.1744V} 1D proper motion dispersion of
the bright ONC stars of 0.7~mas/yr, the probability for \toneC\ to be
in the predicted range is 0.023. Indeed, \citet{1988AJ.....95.1744V} already noted
that \toneC\ has an abnormally large proper motion.

(2) {\it ONC-Frame Proper Motion in Right Ascension ($\mu_{\rm \alpha,ONC}({\rm \theta^1C}) {\rm cos}\delta$):} Similarly, in the frame of the center of mass of the pre-ejection
triple: $\mu_{\rm \alpha,T}({\rm \theta^1C}) {\rm cos}\delta =
-(m_{\rm BN}/m_{\rm \theta^1C}) \mu_{\rm \alpha,T}({\rm BN}){\rm
  cos}\delta \rightarrow -([9.3\pm2.0\,\msun]/[47\pm4\,\msun])
(-6.1\pm1.2)\:{\rm mas\:yr^{-1}} \rightarrow +1.21 \pm 0.36\:{\rm
  mas\:yr^{-1}}$. The predicted value of the ONC-frame motion of
\toneC\ is then $\mu_{\rm \alpha,ONC}({\rm \theta^1C}) {\rm cos}\delta [{\rm predicted}]
=+1.2\pm 0.8 \:{\rm mas\:yr^{-1}}$. The observed
value \citep{1988AJ.....95.1744V} is $\mu_{\rm \alpha,\theta^1C} {\rm
  cos} \delta [{\rm observed}] = +1.4\pm 0.2\:{\rm mas\:yr^{-1}}$ and the probability
that \toneC\ is in the predicted range by chance is 0.27.  

(3) {\it ONC-Frame Radial Velocity ($v_{\rm r,ONC}({\rm \theta^1C})$):} Similarly, the radial recoil in the frame of the pre-ejection
  triple should satisfy: $v_{\rm r,T}({\rm \theta^1C})= -(m_{\rm
    BN}/m_{\rm \theta^1C}) v_{\rm r,T}({\rm BN})\rightarrow
  -([9.3\pm2.0\,\msun]/[47\pm4\,\msun])(+13\pm 1.8\:\kms)
  \rightarrow  -2.57\pm0.69 \:\kms$. The predicted value of the
  ONC-frame motion of \toneC\ is then $v_{\rm r,ONC}({\rm \theta^1C})[{\rm predicted}]
  = -2.6\pm 1.6\:\kms$, with the error range dominated by the
  assumption that the pre-ejection triple had a motion similar to the
  other bright ONC stars \citep{1988AJ.....95.1744V} with $\sigma_{\rm 1D}=1.4\:\kms$. 
  \toneC\ has an observed heliocentric velocity \citep{2009A&A...497..195K} of $+23.6\:\kms$
  i.e. an LSR velocity of
  $5.5\:\kms$, i.e. an ONC frame velocity of $v_{\rm r,ONC}({\rm \theta^1C})[{\rm observed}] = -2.5\:\kms$.
For a Gaussian distribution with $\sigma_{\rm 1D}=1.4\:\kms$,
i.e. based on the proper motion dispersion of bright stars \citep{1988AJ.....95.1744V} the probability of being in the predicted velocity range
by chance is 0.24. 

(4) {\it Mass of Secondary ($m_{\theta^1{\rm C}_2}$):} Given a
\toneC\ primary mass of $38.2\,\msun$, what is the probability of
having a secondary star with mass $\gtrsim m_{\rm BN}$? We estimate
this probability using the low value ($m_{\rm{BN}} = 7.3\,\msun$) of
the luminosity-based mass estimate for BN of $9.3\pm2.0\,\msun$.  If
the secondary star is drawn from a Salpeter power-law mass function,
$dF/dm_*\propto m_*^{-2.35}$, where $F$ is the fraction of the stellar
population, with maximum mass equal to the primary mass and lower mass
limit equal to $1.0\,\msun$ (a relatively top-heavy IMF, with average
mass of $2.8\:M_\odot$, compared to the global ONC IMF, which has a
broad peak around $0.5\:M_\odot$, e.g., \citealt{2002ApJ...573..366M}), then the
probability of having a secondary with mass $>7.3\,\msun$ is
0.061. \citet{1998ApJ...492..540H} find evidence for mass
segregation in the IMF ({\it of primary stars}) in the center of the
ONC, with average stellar mass reaching peak values of $\sim
1.3-2\:M_\odot$ in the vicinity of \toneC\ (excluding \toneC\ from the
average). If we raise the lower limit of the above Salpeter IMF to
$2\,\msun$ (i.e. an average mass of $5.1\:M_\odot$), this raises the
probability of obtaining a sufficiently massive secondary to 0.16 and
we adopt this number as a conservative estimate. We note that none of the
other Trapezium stars has a secondary mass that satisfies this
condition \citep{2004ApJ...607L..47T}.

(5) {\it Ratio of Ejection Kinetic Energy to Binary Total Energy
  ($E_{\rm ratio}$(BN-\toneC)):} From our numerical experiments we find that in order for
\toneC\ to have ejected BN at its observed velocity, $0.23<E_{\rm ratio}[{\rm predicted}]<0.72$.
This range contains about 70\% of the \bnvel\ events.  Recall, $E_{\rm
  ratio}[{\rm observed}]=0.49\pm 0.22$.  Given the primary and
secondary masses of \toneC, what is the probability the total energy
of the binary, $E_{\rm binary}$, falls in the predicted range of $1.4$
to $4.3\times T_{\rm ejection}$, i.e. $(1.1-3.5)\times 10^{47}\:{\rm
  erg}$, simply by chance?  This corresponds to a range of semi-major
axes of 8.5 to 27~AU. If the distribution of $a$ follows $d F_b / d
{\rm log} a = $ constant \citep{2003gmbp.book.....H} from 0.1 to 6300
AU (see \S\ref{sec:method}), then the probability that the \toneC\ binary
falls in this range is 0.10.
If we assume ejection occurred via Case 3, then $0.44<E_{\rm
  ratio}[{\rm predicted}]<0.83$, corresponding to a range of semi-major
axes of 16 to 30~AU and a probability of chance agreement of 0.057.
The upper limit of allowed $a$ may be smaller than 6300~AU for
conditions in the central regions of the ONC. For a stellar density of
$\sim 10^4\:{\rm pc}^{-1}$, the average separation is 6000~AU. Most of
these stars will have masses lower than \toneCtwo or BN. Hence it is not 
likely that these lower mass stars will disrupt 
the relatively more massive BN or \toneCtwo\ stars from being
in a binary with \toneC$_1$. If we adopt an upper limit smaller by a
factor of 2, i.e. $3000$~AU, then the probabilities of the \toneC\ binary falling in the above
expected ranges by chance rises by just 7\%.

%

\begin{figure*}
\begin{center}
\plotone{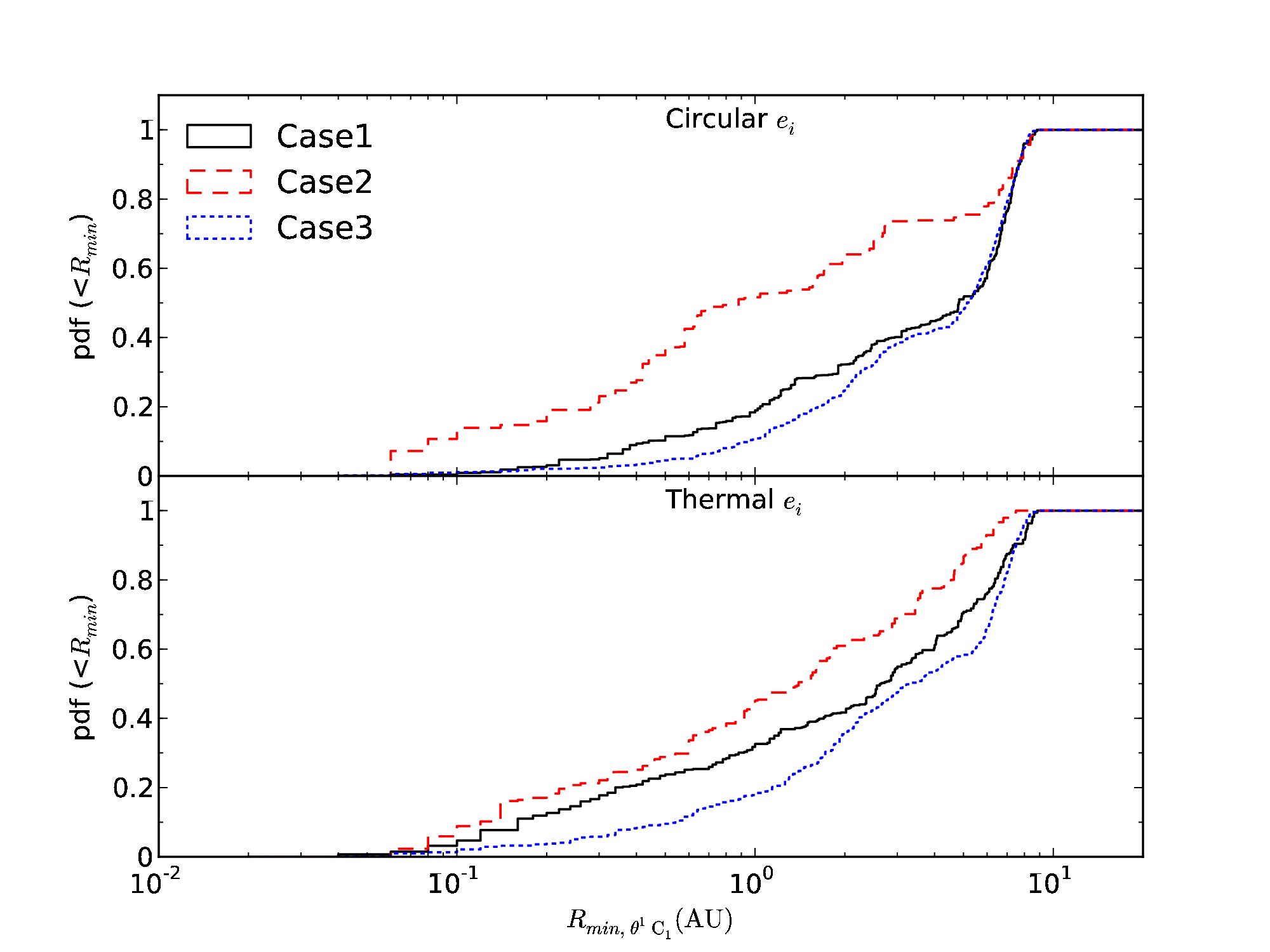}
\caption[Closest approach to \toneCone]{\footnotesize Cumulative histogram 
weighted by respective cross-sections for the closest approach of either BN 
or \toneCtwo\ to \toneCone\ ($R_{min, \theta^1\rm{C}_1}$) during the binary-single 
interactions for all cases and $e_i$-distributions for \truebn\ events.  For all cases 
a significant fraction of interactions results in $R_{min, \theta^1\rm{C}_1} \gtrsim 5\,\au$.  
Especially, for Case 3 and Circular $e_i$-distribution, probability for $R_{min, \theta^1\rm{C}_1} \gtrsim 5\,\au$ 
is about $60\%$.   
}
\label{fig:closeapproach}
\end{center}
\end{figure*}

(6) {\it Eccentricity ($e$(\toneC)):} Our numerical experiments (see
Figure\ 2) show that a very broad range of eccentricities is expected
for the \toneC\ binary if it has ejected BN at the observed velocity
(the average value for all the \bnvel\ outcomes is $e=0.60$ with a
$1\sigma$ range from 0.34 to 0.86; for Case 3 this range is from 0.34
to 0.82). The observed value of $e=0.592\pm0.07$ is consistent with
these expectations, especially being close to the peak of the
distribution resulting from Case 3
(Figure\ \ref{fig:eratio_e}). However, to assess the probability of
this agreement by chance we need to know the eccentricity distribution
of ONC binaries, especially for massive stars. Unfortunately there are
few observational constraints on this eccentricity distribution. If
the binaries have existed long enough to suffer many interactions,
then a thermal distribution, $dF_b/de = 2 e$, is expected, which is
weighted towards high eccentricities. A binary drawn from a thermal
distribution has a 0.62 chance to be in the range $0.34<e<0.86$ (0.56 to
be in the range $0.34<e<0.82$ for Case 3).  Of course, if the actual
distribution of $e$ of ONC binaries is close to circular ($e\simeq
0$), then the probability of chance agreement for the eccentricity of
\toneC\ with the value expected from BN ejection would be very
small. To be conservative, we adopt the probability of 0.62 implied by
a thermal distribution of eccentricities.

(7) {\it Angle between the direction of BN's motion and the angular
  momentum of \toneC\ binary ($\alpha$):} From our numerical
experiments we find that for \bnvel\ events the angle $\alpha$
between $\hat{L}_{\rm{\theta^1C}}$ and $\hat{v}_{\rm{BN}}$ should satisfy 
$54^\circ < \alpha < 126^\circ$.  This range contains about 70\% of all 
\truebn\ events. 
For Case 3, the range is $58^\circ < \alpha < 122^\circ$. 
The observed value of $\alpha$ for the BN-\toneC\ system is $\sim
90^\circ$ and thus contained within this range. If \toneC\ and BN were
unrelated, then the distribution of $\alpha$ should be $0.5\:{\rm
  sin}\:\alpha$ over the range $0^\circ$ to $180^\circ$. Hence the
probability for chance occurrence for $\alpha$ to be within the above
range is 0.59 (0.53 for Case 3).

Combining the above individual probabilities and assuming, reasonably,
that these properties are mutually independent, we find that the total
probability of chance agreement of all 7 properties is small,
$\epsilon = 0.023\times 0.27 \times 0.24 \times 0.16 \times 0.10
\times 0.62 \times 0.59 = 8.7\times 10^{-6}$.  If we assume ejection
happened via Case 3, which affects the last three probabilities, we
obtain a total probability of chance agreement of $\epsilon =
4.0\times 10^{-6}$. The excellent agreement between the observed
BN-\toneC-system properties and the predicted final properties from
our numerical simulations (\S3) together with the low chance agreement
probability ($\epsilon$) of all independent observed properties of the
system strongly supports our proposed scenario that the observed
system resulted from a strong binary-single ejection event involving
\toneCone, \toneCtwo, and BN with a high probability of about $1-\epsilon
\approx 0.99999$.

From the present day velocities and projected distance between the
\toneC\ and BN this binary-single encounter must have happened about
$4500$ years ago at the location shown in Fig.~\ref{fig:onc}. This
scenario also explains the anomalously large proper motion of
\toneC\ \citep{1988AJ.....95.1744V}, which will cause it to leave the
central region of the cluster within $\sim 10^5$ years.

Based on a multi-frequency radial velocity analysis, 
\citet{2010A&A...514A..34L} have suggested \toneC\ may actually harbor an additional
star of $1.0\pm0.16\,\msun$ in a close ($a=0.98\,\au$; $P=61.5\,\rm{day}$),
eccentric ($e=0.49$) orbit around \toneCone. If during the proposed
scattering event that ejected BN, \toneCtwo\ or BN came close to this
inner region then one would expect likely ejection of the solar mass
star. To examine the likelihood of such close interactions, in 
Figure\ \ref{fig:closeapproach} we show the cumulative histogram (weighted by
cross-section) of closest approaches of either \toneCtwo\ or BN to \toneCone\ 
for all \truebn\ events. For example for Case 3 with Circular $e_i$, $\sim 60\%$ 
of the events happen with a closest approach that is
$>5\,\au$. At these distances we would expect the solar mass star to
remain relatively undisturbed in its orbit. Case 1 and especially Case
2 involve somewhat closer approaches, although both of them still have
at least a $20\%$ contribution to events with closest approach
$>5\,\au$. Future confirmation of the reality of the third star in the
\toneC\ system may help us place further constraints on how BN was
ejected from \toneC.

\begin{deluxetable*}{cccc}
\tabletypesize{\footnotesize}
\tablecolumns{4}
\tablewidth{0pt}
\tablecaption{Current properties of \toneC\ that are required if it ejected BN}
\tablehead{\colhead{Property of \toneC} &
           \colhead{Predicted Value} &
           \colhead{Observed Value\tablenotemark{a}} &
           \colhead{Probability of}\\
           \colhead{} &
           \colhead{} &
           \colhead{} &
           \colhead{Chance Agreement}
}
\startdata
Proper Motion in Dec.\tablenotemark{b} ($\mu_{\rm \delta,ONC}$) & $-2.3 \pm 0.9\:{\rm mas\:yr^{-1}}$ & $-1.8\pm 0.2 \:{\rm mas\:yr^{-1}}$ & 0.023\\
Proper Motion in R.A.\tablenotemark{b} ($\mu_{\rm \alpha,ONC}{\rm cos}\delta$) & $+1.2\pm 0.8 \:{\rm mas\:yr^{-1}}$ & $+1.4\pm 0.2\:{\rm mas\:yr^{-1}}$ & 0.27\\
Radial Velocity\tablenotemark{b} ($v_{\rm r,ONC}$) & $-2.6\pm 1.6\:\kms$ & $-2.5\:\kms$ & 0.24 \\
Mass of Secondary ($m_{\theta^1{\rm C}_2}$) & $7.3 - 38.2\,\msun$ & $8.8\pm1.7\,\msun$ & 0.16\tablenotemark{c}\\
Eject. KE to Binary Total E ($E_{\rm ratio}$) & $0.23 - 0.72$ & $0.49\pm 0.22$ & 0.10 [0.057]\tablenotemark{d}\\
Eccentricity ($e$) & $0.34 - 0.86$ & $0.592\pm0.07$ & 0.62\tablenotemark{e}[0.56]\tablenotemark{d}\\
Angle between $\hat{L}_{\rm{\theta^1C}}$ \& $\hat{v}_{\rm{BN}}$ ($\alpha$) & $54^\circ - 126^\circ$ & $\sim 90^\circ$ & 0.59[0.53]\tablenotemark{d} \\
\hline
Combination of 7 Independent Properties & & & $8.7\times 10^{-6}$[$4.0\times10^{-6}$]\tablenotemark{d}\\
\enddata

\tablenotetext{a}{References: \toneC\ proper motions from \citet{1988AJ.....95.1744V}; Other properties from Kraus et\ al. (2009).}
\tablenotetext{b}{ONC-frame}
\tablenotetext{c}{Assumes secondary is drawn from a Salpeter initial mass function (IMF) with lower mass limit of $2\,\msun$. A lower limit of $1\,\msun$ (still a top-heavy IMF) would yield a probability of 0.061.}
\tablenotetext{d}{Number in square brackets assumes ejection via Case 3.}
\tablenotetext{e}{Assumes binary eccentricity is drawn from a thermal distribution, which is the most eccentric distribution that can be expected (requiring the cluster stars to have had a long enough time to interact). Thus this probability should be regarded as a conservative upper limit.}

\label{tab:properties}
\end{deluxetable*}

%
%
\section{Discussion and Summary}
\label{sec:conclude}

We present the following argument in favor of our proposed scenario
that BN was ejected by a binary-single interaction involving
\toneCone, \toneCtwo, and BN in the past.  BN is a runaway star
\citep{1995ApJ...455L.189P,2004ApJ...607L..47T}. 
It had to be launched by an interaction with a multiple system that
has a primary mass greater than BN's mass along its past
trajectory. The most massive binary in the ONC, \toneC, is a system
satisfying this condition, but there are a few other candidates
including, potentially, the massive protostar source I. To test
whether \toneC\ binary ejected BN we consider 7 additional properties,
namely recoil in 3 directions, sufficiently massive secondary, orbital
binding energy, orbital eccentricity and angle between the orbital
angular momentum vector and the direction of BN's velocity. Aided in
part by a large and well-sampled suite of numerical simulations we
show that all of these 7 observed properties of the BN-\toneC\ system
agree well with the properties predicted if BN was ejected by
\toneC. There are two and only two possibilities: 1) \toneC\ has all
these properties {\it by chance}; 2) \toneC\ has acquired these
properties naturally as a result of BN's ejection. We estimate the
probability of chance agreement for all of the above properties to be
low $\epsilon \lesssim 10^{-5}$. Hence, we conclude with about $(1-\epsilon)
\approx (1-10^{-5})$ probability that today's BN-\toneC\ system was
created via a binary-single interaction involving \toneCone,
\toneCtwo, and BN.

A summary of our numerical calculations is as follows. We
performed $\sim 10^7$ numerical simulations that allowed us to properly
sample the multidimensional parameter space effectively
(\S\ref{sec:method}). We used two different $e_i$ distributions as
limiting cases, and $200$ different $a_i$ values from the full range
of possible $a_i$ in ONC from physical considerations for all three
cases that can produce the observed BN-\toneC\ system
(\S\ref{sec:method}). We found that the predicted energies of the
BN-\toneC\ system agree well with the predictions from the scattering
scenario (Figure\ \ref{fig:eratio_e}) for both assumed $e_i$
distributions and all three cases. Furthermore, the predicted
distribution of the angle $\alpha$ between BN's velocity vector
$\vec{v}_{\rm{BN}}$ and \toneC's angular momentum vector
$\vec{L}_{\theta^1\rm{C}}$ was consistent with the observed value of
$\alpha$ (Figure\ \ref{fig:alpha_sigma}). Further calculations of
cross-sections for the \bnvel\ and \truebn\ events constrained a range
of initial binary semimajor axes for each Case (1, 2, 3) that would
produce the observed BN-\toneC\ system followed by a strong
scattering (Figure\ \ref{fig:a_sigma}). For these ranges of $a_i$ ejection of BN in general is the
dominant outcome (apart from weak fly-bys which has a formally
infinite cross-section; Figure\ \ref{fig:branching}).  Our results
indicate that all Cases can contribute to the production of the
observed BN-\toneC\ system. However, an interaction between an initial
binary with members \toneCtwo\ and BN, and a single star \toneCone,
our Case 3 (denoted by \casethree), is favored by about a factor of 2
(Figure\ \ref{fig:b_sigma}).


Ejection of BN from \toneC\ has several important implications for
our understanding of massive star formation in the KL
nebula, where a core of gas appears to be collapsing to form at least one massive star,
thought to be detected in the radio as source {\it I} (Figure\ 1).

If BN was ejected from \toneC, then its passage near source {\it I}
within $\sim0.5''$, i.e. a projected separation of $\sim 200$~AU,
i.e. an expected physical separation
\citep{2008ApJ...685..333G,2011ApJ...728...15G} of $\sim 300$~AU in
the KL nebula is coincidental. Our estimated ejection point
(Figure\ 1) is $50.1''$ from source {\it I}'s present location,
corresponding to 20,800~AU. The ONC-frame radial velocity is about
half of the plane of sky velocity, implying BN also travelled
10,400~AU in the radial direction to reach source {\it I}, for a total
distance of 23,200~AU. The probability to approach within 300~AU of
source {\it I}, ignoring gravitational focussing, is thus $\pi
(300/23,200)^2 /4\pi = 4\times 10^{-5}$. Gravitational focussing by
$\sim 20\,\msun$ of total mass in and around source {\it I}
\citep{2003ApJ...585..850M,1992ApJ...393..225W} boosts the
cross-section by $\sim$14\%, so the probability of approach is $\sim
5\times 10^{-5}$. Thus the interaction of BN with KL is an improbable
event. Of course, the chance of interaction of BN with {\it any}
existing protostar, not necessarily source {\it I}, in the ONC is
larger simply by a factor equal to the number of protostars in this
volume around \toneC. From X-ray observations
\citep{2005ApJS..160..530G} there appear to be at least $\sim$10 such
objects even in just the local vicinity of KL, so the total
probability of an interaction between BN and a protostar can be at
least an order of magnitude higher, but still leaving it as being
highly unlikely. We note that the probability that \toneC\ is
masquerading as the system that ejected BN is even smaller than the
probability of BN interacting with source {\it I}. Furthermore, the
evidence for \toneC\ to have ejected BN is based on 7 independent
lines of evidence and so survived several tests for falsification
(\S\ref{sec:probability}).

A close passage of BN with source {\it I} will have deflected BN's
motion by an angle\\ $7.5^\circ (m_I/20\,\msun)(b/300{\rm
  AU})^{-1}(v_{\rm BN}/30{\rm km\:s^{-1}})^{-2}$ towards source {\it
  I}, where $b$ is the initial impact parameter and $v_{\rm BN}$ is
the velocity of BN relative to source {\it I}. We expect that a
reasonably accurate estimate of the original ONC-frame position angle
(P.A.) of BN's proper motion can be derived by considering the angle
from the estimated position of the dynamical ejection from
\toneC\ (the cross in Figure\ 1) and BN's current position, which is
$-29^\circ.8$.  The current observed P.A. of BN's ONC-frame proper
motion is variously estimated to be
$-27^\circ.5\pm4^\circ$\citep{2008ApJ...685..333G} and
$-18^\circ.8\pm4.6^\circ$\citep{2011ApJ...728...15G}. The average of
these is $-23^\circ.1\pm3^\circ$, suggesting a projected deflection of
$6^\circ.7\pm\sim 3^\circ$ towards source {\it I}. The total true
deflection may be expected to be $\sim \sqrt{2}$ larger, i.e. $\sim
9^\circ.5\pm 3^\circ$, which is consistent with our previous estimate.

Close passage of BN near the accretion disk of source {\it I} about
500 years ago would induce tidal perturbations that enhance the
effective viscosity of the disk and thus its accretion rate
\citep{2004ApJ...607L..47T}.  This can explain the enhanced,
apparently explosive, outflow \citep{1993Natur.363...54A}, which has
observed timescales of $\sim 500-1000$~yr, if the inner region of the
disk with an orbital time $\lesssim 500$~yr has been significantly
perturbed. This corresponds to disk radii of $r\lesssim 180
(m_I/20\,\msun)^{1/3}$~AU, which is consistent with the estimate of
closest approach based on deflection of BN's trajectory.

The estimated current accretion rate to source {\it I} is $\sim
3\times 10^{-3}\,\msun yr^{-1}$ to a $10\,\msun$ protostar (and
somewhat higher rates if the protostar is more massive)
\citep{2010A&A...522A..44T}.  This is about a factor of 10 higher than
expected given the properties of the gas core
\citep{2002Natur.416...59M,2003ApJ...585..850M}.  If constant over the
last $1000$~yr, this would imply a total accreted mass of about
$3\,\msun$, which would be a significant fraction of the original
accretion disk-mass around a $10-20\,\msun$ protostar, since the disk
mass is expected to be limited to $\sim 30\%-100\%$ of the stellar
mass by gravitational torques \citep{2010ApJ...708.1585K}. The mass
launched by a magneto-hydrodynamic outflow during this time is
expected to be $\approx 30\%$ of this amount
\citep{1994ApJ...429..808N}, i.e. about $1\,\msun$. This is consistent
with the mass estimated \citep{1982ApJ...259L..97C} to be in the
inner, ``explosive'' part of the outflow of about $3\,\msun$.

Our results suggest that the formation of a massive star in the KL
nebula, i.e. source {\it I}, has been affected by an external
perturbation of a runaway B star, BN, ejected from a different region
of the cluster, i.e. by \toneC. For BN to be launched so close to a
forming massive protostar does appear to be an intrinsically unlikely
event, but multiple pieces of independent evidence strongly support
this scenario. It also explains source {\it I}'s anomalously high
accretion rate and the unusual, apparently ``explosive'' nature of the
recent outflow from this source. It is possible that a significant
fraction, up to $\sim 1/3$, of the accreted mass has been induced by
this perturbation. In other respects, the Core Accretion model
\citep{2002Natur.416...59M,2003ApJ...585..850M} of massive star
formation provides a reasonable description of the system, e.g., the
presence of a massive core around the protostar and two wide-angle
outflow cavities from which near-IR light emerges (Testi et
al. 2010). The example of Orion BN-KL suggests that, occasionally, in
crowded regions near the center of star clusters, the star formation
model needs to be modified to account for tidal perturbations from
external, passing stars.

Improved observational constraints on the properties of \toneC\,
especially its binary properties and ONC-frame proper motion, and the
mass and current proper motion of BN will place even more stringent
tests on the proposed ejection scenario and also help improve the
dynamical mass constraints on source {\it I} from the deflection of BN.

\acknowledgments We thank John Bally, Paola Caselli, Eric Ford, John Fregeau,
Ciraco Goddi, Lincoln Greenhill, Stefan Kraus, Chris McKee, Francesco Palla, Hagai
Perets, Dick Plambeck and Leonardo Testi for helpful discussions. We thank the anonymous 
referee for a detail review. JCT acknowledges
support from NSF CAREER grant AST-0645412; NASA Astrophysics Theory
and Fundamental Physics grant ATP09-0094; NASA Astrophysics Data
Analysis Program ADAP10-0110 and a Faculty Enhancement Opportunity
grant from the University of Florida.  SC acknowledges support from
the Theory Postdoctoral Fellowship from UF Department of Astronomy and
College of Liberal Arts and Sciences.

\end{document}